\documentclass[aps,twocolumn,amsmath,amssymb,preprintnumbers]{revtex4}
\usepackage{amsmath} \usepackage{amsfonts} \usepackage{amssymb}
\usepackage{wasysym}
\usepackage{bbm}
\usepackage{epsfig}
\usepackage{graphics}
\usepackage{graphicx}
\newcommand{\be}{\begin{equation}}
\newcommand{\ee}{\end{equation}}
\newcommand{\ba}{\begin{eqnarray}}
\newcommand{\ea}{\end{eqnarray}}
\newcommand{\nn}{\nonumber}
\newcommand{\kr}{\rangle}
\newcommand{\kl}{\langle}
\newcommand{\y}{(\tilde y)}
\newcommand{\m}{{^m}}
\newcommand{\n}{{^n}}

\begin{document}

\title[ ]{Emergent gravity in two dimensions}

\author{D. Sexty, C. Wetterich}
\affiliation{Institut  f\"ur Theoretische Physik\\
Universit\"at Heidelberg\\
Philosophenweg 16, D-69120 Heidelberg}

\begin{abstract}
  We explore models with emergent gravity and metric by means of
  numerical simulations. 
 A particular type of 
 two-dimensional non-linear sigma-model is
  regularized and discretized on a quadratic lattice. It is
  characterized by lattice diffeomorphism invariance which ensures
  in the continuum limit the symmetry of general coordinate transformations. 
We observe a
  collective order parameter with properties of a metric, showing
  Minkowski or euclidean signature. The correlation functions of the
  metric reveal an interesting long-distance behavior with power-like
  decay. This universal critical behavior occurs without tuning of
  parameters and thus constitutes an example of ``self-tuned
  criticality'' for this type of sigma-models. We also find a
  non-vanishing expectation value of a ``zweibein'' related to the
  ``internal'' degrees of freedom of the scalar field, again with
  long-range correlations. The metric is well described as a composite
  of the zweibein. A scalar condensate breaks euclidean rotation
  symmetry.
\end{abstract}

\maketitle

\section{Introduction}
\label{Introduction}

There are many attempts to formulate a quantum field theory for
gravity by use of a lattice regularization. 
Regge-Wheeler lattice gravity\cite{RW} employs
the lenghts of edges of simplices as basic degrees of freedom
and therefore uses directly elements of (discrete) geometry.
Different geometrical objects are used in other 
formulations of lattice gravity\cite{AJL,RO}.
Lattice spinor gravity\cite{LSG} is an example where the 
basic degrees of freedom are fermions, while no basic geometrical 
objects are introduced.
An extension of lattice spinor gravity uses in addition
to the spinors a geometrical field, namely a connection\cite{Dia}.

We follow here the approach that the metric is obtained as the
expectation value of a suitable collective field, while geometrical
quantities are not used as fundamental degrees of freedom\cite{SG,Vol}.
 In this
sense gravity and geometry emerge from a model of other,
non-geometrical fields. In our approach, the decisive ingredient is
diffeomorphism symmetry in the continuum limit. In four dimensions,
diffeomorphism symmetry entails under rather general circumstances the
presence of a massless spin-two particle, and therefore of a metric
and the corresponding geometry. We employ a lattice formulation with
the property of lattice diffeomorphism invariance \cite{LDI} of the
action and functional measure. This induces diffeomorphism symmetry
whenever the model exhibits non-trivial long range physics which
allows to formulate a continuum limit.

Lattice diffeomorphism invariant models have been proposed using
fermions as fundamental degrees of freedom - namely lattice spinor
gravity \cite{LSG}. A different approach is formulated as a non-linear
$\sigma$-model \cite{LDI}. This latter approach offers the important
advantage that relatively cheap numerical simulations can be used in
order to compute expectation values and correlation functions of the
collective metric field. The purpose of the present paper is a first
numerical study of such lattice diffeomorphism invariant non-linear
$\sigma$-models.

Our first approach concentrates on two dimensions. Sometimes it is
said that two-dimensional gravity is trivial since it does not exhibit
a propagating degree of freedom. This statement holds, however, only
if the quantum effective action is given by an Einstein-Hilbert term
proportional to the curvature scalar $R$, plus a cosmological
constant. Indeed, in two dimensions $R$ is a topological invariant
which cannot provide a kinetic term for the metric. 
We emphasize that there is no reason
to believe that the quantum effective action for a composite metric
should be purely of the Einstein-Hilbert type. Propagating metric
degrees of freedom become possible for a different form of the 
quantum effective
action. We demonstrate this in appendix \ref{appa} with a rather simple
possible form of a diffeomorphism symmetric effective
action. Similarly, we give in appendix \ref{appb} an example for a
diffeomorphism symmetric effective action for a propagating
zweibein. The effective action for the model investigated in the
present paper turns out to be more complicated than these simple
examples.

In this paper we investigate a non-linear $\sigma$-model with two complex scalar fields $\varphi_i$, with ``flavor'' index $i=1,2$. The constraint 
\be\label{I1}
\sum_i\varphi^*_i\varphi_i=1
\ee
is compatible with an $SO(4)$-flavor symmetry acting on the four real fields $\varphi_{i,R},\varphi_{i,I},\varphi_i=\varphi_{i,R}+i\varphi_{i,I}$. We consider the real action $\big(x^\mu=(x^0,x^1)\big)$
\be\label{I2}
S=\beta\int d^2x\epsilon^{\mu\nu}(\varphi^*_1\partial_\mu\varphi_1-\varphi_1\partial_\mu\varphi^*_1)
(\varphi^*_2\partial_\nu\varphi_2-\varphi_2\partial_\nu\varphi^*_2),
\ee
with $\epsilon^{01}=-\epsilon^{10}=1,~\epsilon^{00}=\epsilon^{11}=0$. For continuous space this action is invariant under general coordinate transformations. No metric is introduced a priori - diffeomorphism symmetry is realized by the particular contraction of two derivatives with the $\epsilon$-tensor. 

We discretise the action (\ref{I2}) on a two dimensional quadratic
lattice by using lattice derivatives and cell
averages as explained in detail in section \ref{Action}. 
We perform a numerical study of this model by Monte Carlo technics.
The continuum diffeomorphism symmetry of the action implies diffeomorphism
symmetry of the discretised lattice action. However, we stress that
the effective action describing the system in the continuum limit 
is not of the simple form (\ref{I2}).

Our main findings are the following:
\begin{itemize}
\item [(i)]
  We identify collective fields that have the transformation
  properties of a metric . They acquire indeed non-vanishing
  expectation values. There are several candidates for collective
  metric fields. For the vacuum we find non-vanishing flat metrics both with
  Minkowski and euclidean signature. The metric turns out to describe
  flat space independently of the parameters of the model, which
  raises interesting questions concerning a self-adjustment of the
  effective two-dimensional cosmological constant to zero.
\item[(ii)] The correlations of the metric fluctuations are long range. They typically show a powerlike decay $\sim r^{-\alpha}$, with $\alpha$ close to two.
\item[(iii)] The geometry differs from flat space space if sources
  coresponding to an energy momentum tensor are introduced.  In the
  linear approxiamtion the perturbation of the metric in response
  to a source is determined by the correlation function. A static
  point source, which mimicks a static massive object, leads to a 
``Newtonian'' potential that decays with the 
inverse of the spatial distance. This is similar to four dimensional gravity,
but quite different from naive dimension estimates.

\item[(iv)] The critical behavior associated to the powerlaw for the correlation functions occurs independently of the detailed values of the parameter $\beta$ characterizing the model. The non-linear $\sigma$-model (\ref{I2}) is an example for self-tuned criticality. 
\item[(v)] We also identify a collective zweibein $e^m_\mu$, where the ``Lorentz index'' $m$ is associated to the flavor degrees of the scalar field. To a good approximation, some of the metric candidates can be described as the usual bilinear of the zweibein. 
\item[(vi)] The correlation functions of the zweibein decay with a power of the distance, similar to the metric. 
\item[(vii)] We identify the exact ground state of the model for $\beta\to\infty$. It consists of ``stripe-configurations'' for the scalar field.
\item[(viii)] The stripes persist for $\beta>\beta_c,\beta_c\approx 4.6$. 
  They also characterize the vacuum state of a corresponding continuum
theory. The scalar order parameters describing the stripes 
are responsible for a spontaneous breaking of euclidean rotation symmetry.
 For this reason we encounter, in case of euclidean 
signature, an unusual
  version of $d=2$ gravity. While diffeomorphism symmetry is expected
  to be realized in the continuum limit this does not hold for rotation
  symmetry, due to the presence of preferred axes.
\item[(ix)] At $\beta_c$ we find a first order transition. The disordered phase for $\beta < \beta_c$ shows no stripes and no expectation value for the zweibein. The correlations are short range in this phase. 
\end{itemize}

Our paper is organized as follows. 
In section \ref{Action} we describe the discretisation 
of our model on a lattice. 
We define scalar order parameters and collective metric fields in 
sect. \ref{Metric and order parameters}.
In section \ref{Long range metric} we show 
that the metric correlation functions have a long range, power 
law behaviour. 
The stripe configurations that dominate the system for large values of 
$\beta$ are discussed in sect. \ref{Stripes}.
We find the exact ``ground state'' of the system for 
$\beta \rightarrow \infty$. It is a stripe configuration 
for the scalar fields.
In section \ref{Zweibein} we define and describe the zweibein
as a bilinear in the scalar fields. The ``internal index'' or 
``Lorentz-index'' of the zweibein is related to the flavor structure for the 
scalars.
In section \ref{Generalized Lorentz transformations} 
we discuss the presence of an approximate Lorentz symmetry of our non-linear
$ \sigma$-model. While the action (\ref{I2}) is Lorentz symmetric, 
the constraint (\ref{I1}) violates Lorentz symmetry.
We describe the metric and zweibein correlation functions
in sections \ref{Correlation functions} and \ref{Zweibein correlations},
respectively. Finally, we conclude in section \ref{Conclusions}.

The results presented in this paper are all obtained by numerical 
simulations.
In parallel, we present in a series of appendices our analytical 
investigations of the possible form of effective 
action for two-dimensional gravity. These analytical considerations 
all apply to the continuum limit and exploit the symmetries of our
model. We proceed on various levels: App.~\ref{appa} discusses 
the metric as a unique degree of freedom, while App.~\ref{appb} uses the 
zweibein. In App.~\ref{Field equations} we investigate the effective action 
for the scalar fields, and App.~\ref{appe} combines scalars and 
the zweibein. Only the effective actions in App.~\ref{Field equations} 
and \ref{appe}
describe parts of our numerical findings in a realistic way. 
Finally, App.~\ref{Zweibein correlations from scalar correlations} 
discusses a possible connection between
the correlation functions for the zweibein and the one 
for the scalars.

\section{Lattice Action}
\label{Action}

We regularize our model on a lattice, $x^0=\tilde z^0\Delta,x^1=\tilde z^1\Delta$, with $\tilde z^\mu$ integers such that the sum $\tilde z^0+\tilde z^1$ is odd. This corresponds to a square lattice with lattice distance $\sqrt{2}\Delta$ and nearest neighbors in the diagonal directions $\sim x^0\pm x^1$.  For the corresponding ``diagonal lattice vectors'' the $\tilde z$-coordinates are 
\be\label{S3}
E_0=(1,1)~,~E_1=(-1,1).
\ee
We define cells located on the sites of the dual lattice at $x^\mu=\tilde y^\mu\Delta$, with $\tilde y^\mu$ integer and $\tilde y^0+\tilde y^1$ even. Each cell consists of four lattice points with lattice coordinates $\tilde z^\mu=\tilde y^\mu\pm (\tilde v_\nu)^\mu$, with two unit vectors $\tilde v_\nu$ obeying $(\tilde v_\nu)^\mu=\delta^\mu_\nu$. The lattice derivative at $\tilde y$ is given by
\be\label{I3}
\partial_\mu\varphi(\tilde y)=\frac{1}{2\Delta}\big[\varphi(\tilde y+v_\mu)-\varphi(\tilde y-v_\mu)\big],
\ee
and cell averages obey
\be\label{I4}
\bar\varphi(\tilde y)=\frac14
\big[\varphi(\tilde y+v_0)+\varphi(\tilde y-v_0)+\varphi(\tilde y+v_1)+\varphi(\tilde y-v_1)\big].
\ee
The lattice action replaces in eq. \eqref{I2} the derivatives by lattice derivatives, the fields without derivatives by the cell averages, and $\int d^2x\to \sum_{\tilde y} V(\tilde y)$ with cell volume $V(\tilde y)=2\Delta^2$. The distance between neighboring cells is $\sqrt{2}\Delta$. 

Most important for our context, the action is lattice diffeomorphism invariant. Indeed we can change the positioning of the lattice points $x^\mu(\tilde z)$ from the regular lattice $x^\mu(\tilde z)=\Delta\tilde z^\mu$ to an arbitrary neighboring lattice $x^\mu(\tilde z)=\Delta\tilde z^\mu+\xi^\mu(\tilde z)$. Here the cartesian coordinates $x^\mu$ parametrize a manifold which is some region in $\mathbbm{R}^2$. For general positions of the lattice points on the manifold the lattice derivatives read \cite{LDI}
\ba\label{I5}
&&\partial_0\varphi(\tilde y)=\frac{1}{2V(\tilde y)}
\Big\{\big(x^1(\tilde y+v_1)-x^1(\tilde y-v_1)\big)\nn\\
&&\hspace{1.5cm}\big(\varphi(\tilde y+v_0)-\varphi(\tilde y-v_0)\big)\nn\\
&&~-\big(x^1(\tilde y+v_0)-x^1(\tilde y-v_0)\big)
\big(\varphi(\tilde y+v_1)-\varphi(\tilde y-v_1)\big)\Big\},\nn\\
&&\partial_1\varphi(\tilde y)=\frac{1}{2V(\tilde y)}
\Big\{\big(x^0(\tilde y+v_0)-x^0(\tilde y-v_0)\big)\nn\\
&&\hspace{1.5cm}\big(\varphi(\tilde y+v_1)-\varphi(\tilde y-v_1)\big)\\
&&~-\big(x^0(\tilde y+v_1)-x^0(\tilde y-v_1)\big(\varphi(\tilde y+v_0)-\varphi(\tilde y-v_0)\big)\Big\},\nn
\ea
and the cell volume becomes
\be\label{I6}
V(\tilde y)=\frac12\epsilon_{\mu\nu}\big(x^\mu(\tilde y+v_0)-x^\mu(\tilde y-v_0)\big)
\big(x^\nu(\tilde y+v_1)-x^\nu(\tilde y-v_1)\big).
\ee
The cell averages \eqref{I4} remain the same. The change of each term in the action due to the change of the lattice derivatives is precisely canceled by the change of the volume factor in $\int d^2x=\sum_{\tilde y}V(\tilde y)$. The expression of the lattice action in terms of lattice derivatives and cell averages is therefore independent of the positioning of the lattice points. This crucial property is due to the particular contraction of the lattice derivatives with the $\epsilon$-tensor. It guarantees the standard diffeomorphism symmetry of the quantum effective action in the continuum limit \cite{LDI}. 

We will impose periodic boundary conditions on a square lattice
defined by the $E_0$ and $E_1$ vectors. We have checked by performing 
calculations on a square lattice defined by the $x_0$ and $x_1$ axes, that 
the boundary conditions do not play an important role in determining the 
phase structure of the theory. The functional integral 
\be\label{I7}
Z=\int {\cal D}\varphi e^{-S}
\ee
involves for every lattice point $\tilde z$ the standard $SO(4)$ invariant measure on the sphere $S^3$ in the field space. The functional integral is finite for a finite number of lattice points. Our model is therefore mathematically well defined and a candidate for regularized quantum gravity.

We work with a fixed positioning of the lattice points on the regular quadratic lattice as described above. For general relativity, this corresponds to a fixed choice of coordinates. 

We obtain the numerical results presented in this paper by Monte Carlo
simulations using the standard Metropolis algorithm 
on $64^2$ or $256^2$ lattices. We usually start the thermalization process 
in the stripe phase (see sect. \ref{Stripes}). We have checked that the system
eventually ends up in the stripe phase even if we start the thermalization 
in the disordered phase (for the lattice sizes employed here). In this case 
the thermalization typically takes much longer, as the system first 
breaks up into domains with different orientations of stripes, which 
slowly equilibrate into one coherent domain.

\section{Order parameters and metrics}
\label{Metric and order parameters}

The continuum action \eqref{I2} is invariant under translations and
rotations, as well as with respect to a parity-type discrete
transformation $P$: $x^1\to -x^1,\varphi_1\leftrightarrow \varphi_2$
or time reversal $T:~x^0\to-
x^0,~\varphi_1\leftrightarrow\varphi_2$. Similarly, $S$ does not
change under diagonal reflections $D_\pm:~x^0\leftrightarrow \pm
x^1,~\varphi_1\leftrightarrow\varphi_2$. Another discrete symmetry is
charge conjugation which is realized by complex conjugation,
$C:\varphi_i\to\varphi^*_i$. Furthermore, the action conserves a
global continuous flavor symmetry with abelian gauge group $U(1)\times
U(1)$, corresponding to separate phase rotations for $\varphi_1$ and
$\varphi_2$. Since the action changes sign under
$\varphi_1\leftrightarrow \varphi_2$ we can restrict the discussion to
positive $\beta$. We also observe that $S$ changes sign if only one of
the fields is replaced by its complex conjugate, say
$\varphi_1\to\varphi^*_1,\varphi_2\to\varphi_2$. For the reflections
$P,T,D_\pm$ we can therefore replace the accompanying reflection
$\varphi_1\leftrightarrow\varphi_2$ by
$\varphi_1\leftrightarrow\varphi^*_1$. Combined flavor reflections, as
$\varphi_1\to \varphi_2,\varphi_2\to\varphi_1^*$, leave the action
invariant. 

The lattice action is invariant under diagonal translations
of $\sqrt{2}\Delta$, and therefore also under translations in the
$x^0$ or $x^1$ directions by $2\Delta$. It is preserved by
$\pi/2$-rotations and shares the same discrete symmetries as discussed
before for continuous space, as well as the continuous $U(1)\times
U(1)$ flavor symmetry.

\begin{figure}[htb]
\begin{center}
\includegraphics[width=0.45\textwidth]{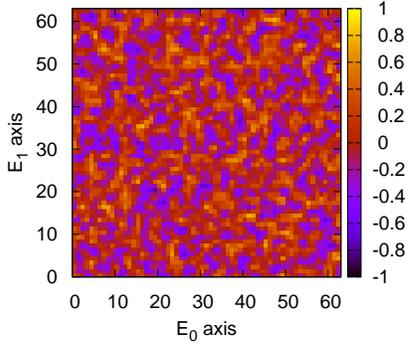}
\includegraphics[width=0.45\textwidth]{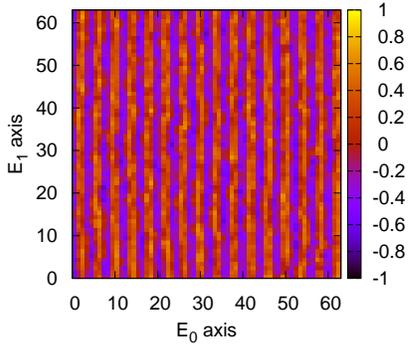}
\caption{The real part of $\varphi_1$ for a typical configuration at 
$ \beta=4< \beta_c $ (top) and $\beta=5 > \beta_c $ (bottom) on a $64^2$ lattice. Units along the axes are $ \sqrt{2} \Delta$. }
\label{qgtd:fig2}
\end{center}
\end{figure}

\begin{figure}[htb]
\begin{center}
\includegraphics[width=0.45\textwidth]{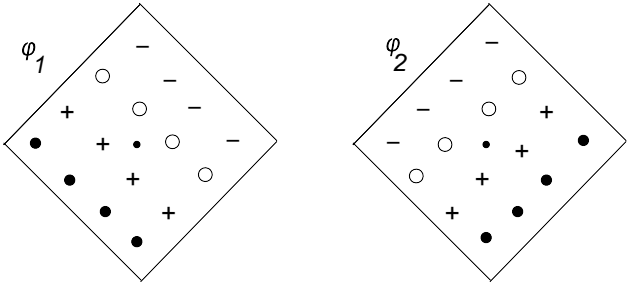}
\caption{Phases of fields in a supercell, $1$ for $+$,  $-1$ for  $-$, $i$ for $\CIRCLE$, $-i$ for $\Circle$. The small dot in the center is the location of the supercell.}
\label{qgtd:fig3}
\end{center}
\end{figure}

We find that for large enough $\beta$ several of the discrete
symmetries are spontaneously broken. Indeed, the characteristic
configurations of the scalar fields change qualitatively at $\beta_c
\approx 4.6 $, as demonstrated in Fig. \ref{qgtd:fig2}. For
$\beta>\beta_c$ one observes an order in stripes which disappears for
$\beta<\beta_c$. For the ``stripe phase'' $(\beta>\beta_c)$ we can
define order parameters $s^\pm_i$ by defining ``supercells'' with
sixteen lattice points. For $s^+_i$ the
field values $\varphi_i$ within a given supercell are summed with
phases according to the left part of Fig. \ref{qgtd:fig3}, while for
$s^-_i$ we sum with the complex conjugate of the phase factors,
\ba \label{opdef}
s_i^+ = {1 \over L^2 } 
\sum\limits_{\tilde z } e ^ { i \eta ( \tilde z ) }
\varphi_i (\tilde z ),  \nonumber \\
s_i^- = {1 \over L^2 } 
\sum\limits_{\tilde z } e ^ { i \eta^* ( \tilde z ) }
\varphi_i (\tilde z ).
\ea
(More precisely, the phases $\exp ( i \eta(\tilde z)) $
depend on the coordinates
within the supercell, as defined by the left panel of Fig.~\ref{qgtd:fig3},
and $L$ is the number of lattice points along one
 (diagonal) direction.)
We will show in sect. \ref{Stripes} that there are four different 
classes of stripe configurations which can not be rotated into 
each other using the internal $ U(1) \times U(1) $ symmetry.
Each of the four order parameters $ s^a_i$ with $i=1,2$ and $a=\pm$ signals 
the realization of 
one of the four equivalent classes of stripe configurations. 

\begin{figure}[htb]
\begin{center}
\includegraphics[width=0.4\textwidth]{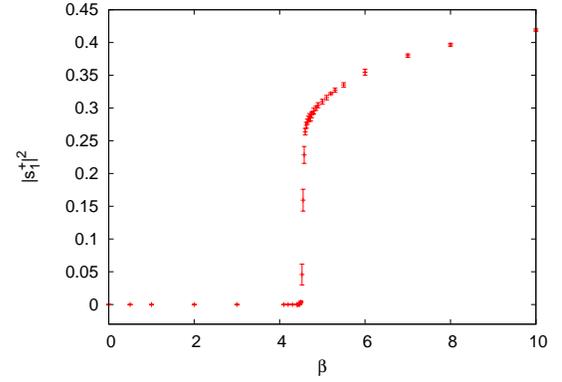}
\caption{The order parameter $ |s_1^+|^2 $ as a function of $\beta$. 
The thermalization of the system was started from the striped phase
signalled by $s_1^+$.}
\label{qgtd:figop}
\end{center}
\end{figure}

We show a characteristic order parameter as a function of $\beta$ in
Fig. \ref{qgtd:figop}. In the presence of such order the lattice
translation symmetry is partially broken - only translations by
$4\sqrt{2}\Delta$ in the diagonal directions or by $8\Delta$ in the
$x^0$ or $x^1$ directions leave the equilibrium state invariant. Also
the discrete reflection symmetries and the symmetry of $\pi/2$-lattice
rotations are broken spontaneously by the direction of the stripes.
The symmetry of the continuous flavor transformations gets broken
spontaneously as well and one may expect Goldstone-type
excitations. All these symmetries are preserved in the disordered
phase for $\beta<\beta_c$. The discontinuity in the scalar order
parameter visible in Fig. \ref{qgtd:figop} indicates a first order
phase transition. This will be confirmed by jump in other expectation
values.

For $\beta>\beta_c$ we will see in sect.~\ref{Stripes} that lattice
translations, reflections and rotations leave the equilibrium state
invariant if they are combined with appropriate phase (or flavor)
rotations. It is therefore perhaps more appropriate to associate the
stripe phase with a spontaneous breaking of the $U(1)\times
U(1)$-flavor symmetry. (We discuss compatibility with the
Mermin-Wagner theorem at the end of sect. \ref{Stripes}.) Observables
which are invariant under flavor rotations will be invariant under
translations by $E_0$ or $E_1$.

We are interested in geometry and therefore look for expectation values of observables that can play the role of a metric.  In the continuum limit such observables should transform as second rank symmetric tensors. We will find several natural candidates for composite metric observables, with euclidean or Minkowski signature of the expectation values. They are typically invariant under flavor rotations. 

Metric tensors can be constructed from derivatives of $\varphi_i$. Many possibilities exist for constructing objects that transform as a symmetric second rank tensors. In the continuum limit $\varphi$ transforms under diffeomorphisms as a scalar and $\partial_\mu\varphi$ as a covariant vector. The symmetric product of two vectors transforms therefore as symmetric second rank covariant tensor, which is precisely the transformation property of the metric. The lattice analogon replaces the derivatives by lattice derivatives \eqref{I3} and the fields without derivatives by the cell averages \eqref{I4}. We discuss here four examples that are all invariant under the flavor symmetry of continuous phase rotations of $\varphi_1$ and $\varphi_2$. 

\begin{figure}[htb]
\begin{center}
\includegraphics[width=0.4\textwidth]{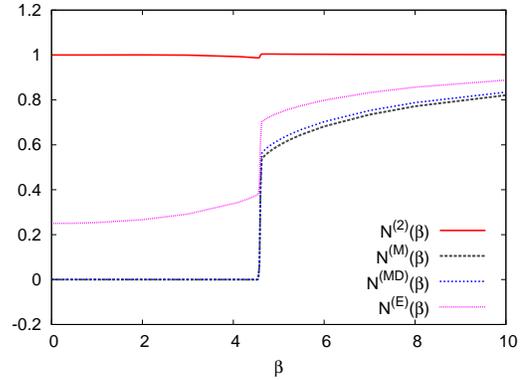}
\caption{Proportionality factor $N(\beta)$ 
of the metric expectation values defined in 
eqs. \eqref{I8}, \eqref{9AA}, \eqref{9B} 
and \eqref{9D}, as a function of $\beta$.  }
\label{qgtd:fig1}
\end{center}
\end{figure}

Our first candidate for a metric reads
\be\label{I8}
\tilde g^{(2)}_{\mu\nu}(\tilde y)= 2\Delta^2 Re\Big(\sum_i\partial_\mu\varphi^*_i(\tilde y)\partial_\nu\varphi_i(\tilde y)\Big),
\ee
with lattice derivatives $\partial_\mu\varphi(\tilde y)$ given by eq. \eqref{I3}. On a $L\times L$ lattice we find that the expectation value is invariant under lattice translations,
\be\label{I9}
g^{(2)}_{\mu\nu}(\tilde y)=\kl \tilde g^{(2)}_{\mu\nu}(\tilde y)\kr=N^{(2)}(\beta)\delta_{\mu\nu},
\ee
with $N^{(2)}(\beta)$ shown in Fig. \ref{qgtd:fig1}. The corresponding geometry is euclidean flat space for all $\beta$. (The normalization factor $N^{(2)}(\beta)$ can be absorbed by a rescaling of the metric or the coordinates.) We observe a discontinuity of $N^{(2)}(\beta)$ at $\beta_c$. The particular ground state $g_{\mu\nu}\sim\delta_{\mu\nu}$ preserves the lattice rotations and reflections.

A second candidate is
\be\label{9AA}
\tilde g^{(M)}_{\mu\nu}=8\Delta^2 Re(i\varphi^*_1\partial_\mu\varphi_1)Re(i\varphi^*_2\partial_\nu\varphi_2)
+(\varphi_1\leftrightarrow\varphi_2).
\ee
For this ``Minkowski metric'' we find that the expectation value 
\be\label{9A}
g^{(M)}_{\mu\nu}=\kl \tilde g^{(M)}_{\mu\nu}\kr = N^{(M)}(\beta)\eta_{\mu\nu}
\ee
has Minkowski signature $\eta_{\mu\nu}=diag(-1,1)$. Again, $N^{(M)}(\beta)$ is plotted in Fig. \ref{qgtd:fig1} and shows a discontinuity at $\beta_c$. The expectation value \eqref{9A} is the same for all lattice sites. It is invariant under (euclidean) rotations by $\pi/2$ and under the discrete symmetries $P$ and $T$. 

A third metric can be written as 
\be\label{9B}
\tilde g^{(MD)}_{\mu\nu}=8\Delta^2 \{ Re(i\varphi^*_1\partial_\mu \varphi_1)Re(i\varphi^*_1\partial_\nu\varphi_1)
-(\varphi_1\leftrightarrow\varphi_2)\}.
\ee
The expectation value $g^{(MD)}_{\mu\nu}=\kl\tilde g^{(MD)}_{\mu\nu}\kr$ is found as
\be\label{9C}
g^{(MD)}_{00}=g^{(MD)}_{11}=0~,~g^{(MD)}_{01}=N^{(MD)}(\beta).
\ee
Again, this corresponds to a Minkowski signature since $\det(g^{(MD)}_{\mu\nu})=-\big(N^{(MD)}(\beta)\big)^2$. In the continuum, the metric \eqref{9C} obtains from the metric \eqref{9A} by a euclidean rotation of the coordinate axes by $\pi/4$. For the metric \eqref{9C} the time-  and space- axes are given by the diagonal axes in the directions of $E_0=(1,1)$ and $E_1=(-1,1)$. Finally, we investigate a fourth metric
\be\label{9D}
\tilde g^{(E)}_{\mu\nu}=
8\Delta^2 Re(i\varphi^*_1\partial_\mu\varphi_1)Re(i\varphi^*_1\partial_\nu\varphi_1)
+(\varphi_1\leftrightarrow\varphi_2).
\ee
The expectation value corresponds again to euclidean flat space 
\be\label{12A}
g^{(E)}_{\mu\nu}=\kl \tilde g^{(E)}_{\mu\nu}\kr=N^{(E)}(\beta)\delta_{\mu\nu}. 
\ee

It is remarkable that, according to the precise definition of the
metric observable, one finds expectation values that correspond to
flat space, either with a euclidean signature or a Minkowski
signature. The signature apparently depends on the flavor structure of
the metric observable. We will gain later a better understanding of
this issue. The dependence of the metric observable on $\beta$ shown
in Fig. \ref{qgtd:fig1} conforms the discontinuity at the critical
value $\beta_c$ that signals the presence of a first order transition.

\section{Long range metric correlations}
\label{Long range metric}

\begin{figure}[htb]
\begin{center}
\includegraphics[width=0.45\textwidth]{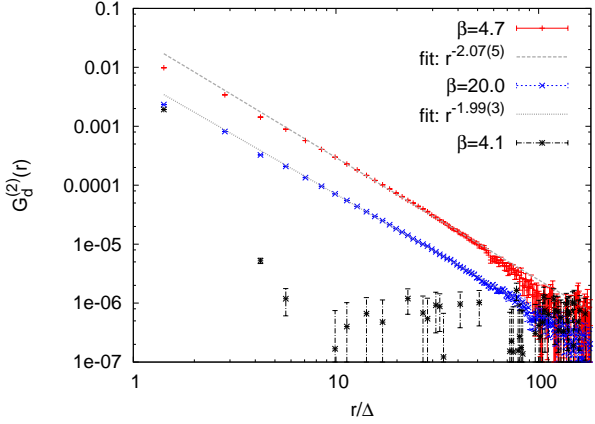}
\caption{Euclidean metric correlator $G^{(2)}_d(r)$,
 defined in eq. (\ref{F2}), as a function 
of distance along the $E_0$ axis, for different couplings $\beta$.}
\label{firstpower}
\end{center}
\end{figure}

In the stripe phase for $\beta>\beta_c$ we observe the presence of long range correlations for fluctuations of the metric field. They decay with a power law. We define the fluctuations 
\be\label{F1}
h^{(A)}_{\mu\nu}=\tilde g^{(A)}_{\mu\nu}-g^{(A)}_{\mu\nu}=\tilde g^{(A)}_{\mu\nu}-
\kl \tilde g^{(A)}_{\mu\nu}\kr
\ee
for the different metric fields \eqref{I8}, \eqref{9AA}, \eqref{9B}, \eqref{9D}, labeled by $(A)$. As an example, we consider the correlation function
\be\label{F2}
G^{(2)}_d(r)=\frac14
\left\kl \big(h^{(2)}_{00}(x)-h^{(2)}_{11}(x)\big)
\big(h^{(2)}_{00}(y)-h^{(2)}_{11}(y)\big) \right\kr
\ee
for $r=|x-y|$. In Fig.~\ref{firstpower} we plot $G^{(2)}_d(r)$ on the diagonal axis $\sim E_0$ for three values of $\beta$. For $\beta>\beta_c$ we clearly see a power law decay. A fit for the range $ 5 \Delta \leq r \leq 50 \Delta$ yields 
\be\label{F3}
G^{(2)}_d(r)=c_dr^{-\alpha_d}
\ee
with 
\ba\label{F4}
\alpha_d&=2.07 \pm 0.05 &~,~c_d=0.031\ \  \text{ for } \beta=4.7 \nn\\
\alpha_d&=1.99 \pm 0.03 &~,~c_d=0.0069\ \text{ for } \beta=20.
\ea

It is remarkable that the powerlaw decay of the correlation function occurs for all values $\beta>\beta_c$. No tuning of a parameter, as common for critical behavior at a second phase transition, is necessary. We have excluded from the fit the points with $r<5\Delta$ since lattice details are relevant in this range - a possible universal behavior will only be found at distances $r$ sufficiently large compared to the lattice distance. For $r>50\Delta$ the correlation becomes very small and the uncertainties large. Also finite volume effects may start to play a role.

On the other hand, we observe in the symmetric phase for $\beta<\beta_c$ a much faster decay of the metric correlation. There are only few points before the correlation gets very small and the uncertainties large. The decay is compatible with an exponential decay or a power law with large negative exponent. Universal long distance behavior seems not to be realized in the symmetric phase, even for $\beta$ close to $\beta_c$.

\begin{figure}[htb]
\begin{center}
\includegraphics[width=0.45\textwidth]{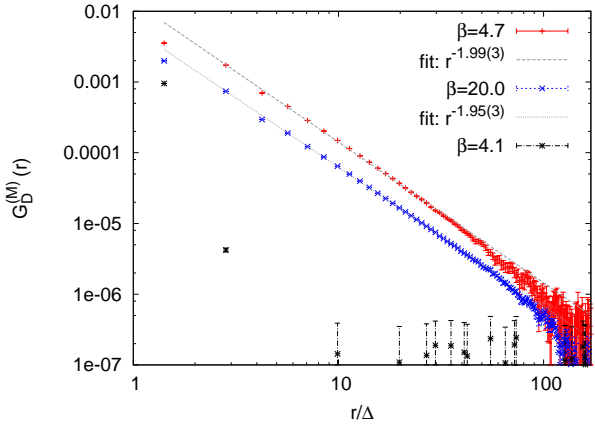}
\caption{Minkowski metric correlator $G^{(M)}_D(r)$,
 defined in eq. (\ref{F2_m}) as a function 
of distance along the $E_0$ axis, for different couplings $\beta$.}
\label{firstpower_m}
\end{center}
\end{figure}

The correlation function for the fluctuations $h^{(M)}_{\mu\nu}$ of the Minkowski metric also shows a long range behavior for $\beta>\beta_c$. 
In Fig.~\ref{firstpower_m} we display $G^{(M)}_D(r)$, defined by
\be\label{F2_m}
G^{(M)}_D(r)=\frac14
\left\kl \big(h^{(M)}_{00}(x)+h^{(M)}_{11}(x)\big)
\big(h^{(M)}_{00}(y)+h^{(M)}_{11}(y)\big) \right\kr
\ee
for $r=|x-y|$ again on the diagonal axis $\sim E_0$. 
This correlation function shows 
a power law decay similar to the euclidean metric $g^{(2)}_{\mu\nu}$, 
with a decay exponent close to $\alpha=2$.
In the following we will concentrate on the stripe phase for $\beta>\beta_c$. Our aim is an understanding of the nature of the observed long range correlations for the metric observables. Further aspects of the correlation
function for the metric will be discussed in sect. \ref{Correlation functions}.

\section{Stripes}
\label{Stripes}

For large values of $\beta$ the leading configurations are stripes. A characteristic stripe configuration reads
\be\label{S1}
\varphi_1(\tilde z)=\frac{1}{\sqrt{2}}
e^{i\alpha_1(\tilde z)}~,~\varphi_2(\tilde z)=\frac{1}{\sqrt{2}}e^{i\alpha_2(\tilde z)},
\ee
with phases 
\ba\label{S2}
\alpha_1\big(\tilde y+(2m_1-1)v_0+n_1E_1\big)&=&-\frac{m_1\pi}{2},\nn\\
\alpha_2\big(\tilde y+(2m_2-1)v_1+n_2E_0\big)&=&-\frac{m_2\pi}{2}.
\ea
Here $m_i$ and $n_i$ are arbitrary integers and $E_0,E_1$ are given by eq. \eqref{S3}. For $\varphi_1$ the stripe is in the diagonal $E_1$-direction, with $\varphi_1(\tilde z+nE_1)=\varphi_1(\tilde z)$, while $\varphi_2$ it is orthogonal to it in the diagonal $E_0$-direction, with $\varphi_2(\tilde z+nE_0)=\varphi_2(\tilde z)$. A typical realistic stripe configuration with fluctuations is shown in the right part of Fig. \ref{qgtd:fig2}. While $\varphi_1$ is invariant under $E_1$-translations, it is invariant under translations in the $E_0$-direction only by four units, $\varphi_1(\tilde z+4nE_0)=\varphi_1(\tilde z)$. Similarly, $\varphi_2$ is invariant under $E_1$-translations by four units, $\varphi_2(\tilde z+4nE_1)=\varphi_2(\tilde z)$. As a whole, the stripe configuration \eqref{S1}, \eqref{S2} is therefore invariant under translations in the $E_0$- and $E_1$-directions by four units. This results in translation invariance in the $x^0$- and $x^1$-directions by $8\Delta$. 

A given stripe configuration is not invariant under the continuous phase rotations of the $U(1)\times U(1)$-flavor symmetry. The phase changes can be used, however, to establish a symmetry of translations in the $E_0$- and $E_1$-directions by one unit combined with an appropriate phase rotation. Indeed, the combined diagonal translations,
\be\label{S4}
t_0:\quad \varphi'_1(\tilde z)=-i\varphi_1(\tilde z-E_0)~,~\varphi'_2(\tilde z)=\varphi_2(\tilde z-E_0),
\ee
and 
\be\label{S5}
t_1:\quad \varphi'_1(\tilde z)=\varphi_1(\tilde z-E_1)~,~\varphi'_2(\tilde z)=-i\varphi_2(\tilde z-E_1),
\ee
leave the stripe configuration \eqref{S1}, \eqref{S2} invariant. By virtue of these combined translations it is often sufficient to evaluate the value that a cell-observable takes for the stripe configuration only for the cell at $\tilde y=0$. The value for the neighboring cell configuration at $\tilde y=E_0$ obtains from the value of the observable for the cell $\tilde y=0$ by multiplying each factor $\varphi_1$ by a phase factor $-i$. Similarly, the value in the neighboring cell at $\tilde y=E_1$ is found by multiplying each factor $\varphi_2$ by a phase $-i$. This can be continued for all cells. In particular, all cell-observables that are invariant under $U(1)\times U(1)$ flavor transformations take for the stripe configuration the same value in each cell.

We first evaluate for the stripe configuration the value of the cell action
\ba\label{S6}
{\cal L}(\tilde y)&=&\epsilon^{\mu\nu}\big[\bar\varphi^*_1(\tilde y)\partial_\mu\varphi_1(\tilde y)-\bar\varphi_1(\tilde y)\partial_\mu\varphi^*_1(\tilde y)\big]\nn\\
&&\times \big[\bar\varphi^*_2(\tilde y)\partial_\nu\varphi_2(\tilde y)-
\bar\varphi_2(\tilde y)\partial_\nu\varphi^*_2(\tilde y)\big], \nn \\
S&= &2 \Delta^2 \beta \sum_{\tilde y} {\cal L} (\tilde y).
\ea
This cell observable is independent of flavor-phase rotations and takes therefore the same values in all cells. For $\tilde y=0$ one finds for the stripe configuration \eqref{S1}, \eqref{S2} the following values
\ba\label{S7}
\bar\varphi_1&=&\frac{1}{2\sqrt{2}}(1-i)~,~\bar\varphi_2=\frac{1}{2\sqrt{2}}(1-i),\nn\\
\partial_0\varphi_1&=&-\frac{1}{2\sqrt{2}\Delta}(1+i)~,~\partial_1\varphi_1=-\frac{1}{2\sqrt{2}\Delta}(1+i),\nn\\
\partial_0\varphi_2&=&\frac{1}{2\sqrt{2}\Delta}
(1+i)~,~\partial_1\varphi_2=-\frac{1}{2\sqrt{2}\Delta}(1+i).
\ea
This yields
\ba\label{S8}
\bar\varphi^*_1\partial_0\varphi_1=-\frac{i}{4\Delta}&,&\bar\varphi^*_1\partial_1\varphi_1=-\frac{i}{4\Delta},\nn\\
\bar\varphi^*_2\partial_0\varphi_2=\frac{i}{4\Delta}&,&\bar\varphi^*_2\partial_1\varphi_2=-\frac{i}{4\Delta},
\ea
and therefore
\be\label{S9}
{\cal L}(\tilde y)=-\frac{1}{2\Delta^2}.
\ee
We will see below that the stripe configuration \eqref{S1}, \eqref{S2} minimizes the action \eqref{I2}. (The action per lattice point takes the value $-\beta$.) The ground state for $\beta\to\infty$ is therefore given by the stripe \eqref{S1}, \eqref{S2} or one of the equivalent configurations that we will discuss below. 

The metric bilinear \eqref{I8} is again a cell observable that does not depend on the flavor phases and therefore takes the same value for all cells. For the stripe configuration one finds from eq. \eqref{S7}
\be\label{S10}
\tilde g^{(2)}_{00}(\tilde y)=\tilde g^{(2)}_{11}(\tilde y)=1~,~\tilde g^{(2)}_{01}(\tilde y)=0. 
\ee
This agrees with the asymptotic value $N^{(2)}(\beta\to\infty)=1$ that can be seen in Fig. \ref{qgtd:fig1}.

We next investigate the discrete lattice reflections which leave the stripe configuration invariant. A time reflection at the axis $\tilde z^0=M$ leaves the stripe invariant if it is accompanied by $\varphi_1\leftrightarrow\varphi_2$ and an appropriate global phase change 
\ba\label{S11}
T_M:&& \varphi'_1(\tilde z^0-M,\tilde z^1)= e^{i\alpha_M}\varphi_2\big(-(\tilde z_0-M),\tilde z_1\big),\nn\\
&&\varphi'_2(\tilde z^0-M,\tilde z_1)=e^{-i\alpha_M}\varphi_1\big(-(\tilde z_0-M),\tilde z_1\big),\nn\\
\ea
where
\be\label{S12}
\alpha_M=-\frac{M\pi}{2}.
\ee
Similarly, a parity type invariance of the stripe configuration obtains by combining the reflection at the axis $\tilde z^1=N$ with an appropriate phase shift
\ba\label{S13}
P_N:&&\varphi'_1(\tilde z^0,\tilde z^1-N)=e^{i\alpha_N}\varphi^*_2\big (\tilde z^0,-(\tilde z^1-N)\big),\nn\\
&&\varphi'_2(\tilde z^0,\tilde z^1-N)=e^{i\alpha_N}\varphi^*_1\big (\tilde z^0,-(\tilde z^1-N)\big),\nn\\
&&\alpha_N=-\frac{(N+1)\pi}{2}.
\ea
Since the symmetry $P_N$ in eq. \eqref{S13} also involves the complex conjugation of the fields $\varphi_i$ it may be associated with a CP transformation. Both symmetries $P_N$ and $T_N$ are symmetries of the action. For a ground state preserving these symmetries an observable and its associated reflected observable must have the same expectation value. 

In particular, we  may consider observables that are invariant under the transformations $\varphi_1\leftrightarrow\varphi_2,\varphi_i\leftrightarrow \varphi^*_i$, as well as under flavor phase rotations. Then the expectation values of observables for which only the coordinates are reflected must be the same as the ones for the original observables. The metric observable \eqref{I8} is of this type. The symmetries $T_M$ and $P_N$ imply $g^{(2)}_{01}(\tilde y)=0$ since $g^{(2)}_{01}$ is odd under the reflection of one coordinate. 

For the diagonal reflections $D_\pm$ we restrict the discussion to reflections at axes through the origin $\tilde y_B=(0,0)$. Reflections at shifted axes can be obtained by a combination of those ``basic reflections'' with the translations $t_0$ and $t_1$ given by eqs. \eqref{S4}, \eqref{S5}. (Also the reflections $T_M$ and $P_N$ can be related in this way to suitable basis reflections at axes through the origin.) The stripe \eqref{S1}, \eqref{S2} is indeed invariant under the diagonal reflections
\ba\label{S17}
\tilde D_+:&& \varphi'_1(\tilde y)=\varphi_1(D_+\tilde y)~,~\varphi'_2(\tilde y)=-i\varphi^*_2(D_+\tilde y),\nn\\
\tilde D_-:&& \varphi'_1(\tilde y)=-i\varphi^*_1(D_-\tilde y)~,~\varphi'_2(\tilde y)=\varphi_2(D_-\tilde y),
\ea
with
\be\label{S18}
D_+\tilde y=(\tilde y^1,\tilde y^0)~,~D_-\tilde y=(-\tilde y^1,-\tilde y^0).
\ee
The symmetries $\tilde D_+$ and $\tilde D_-$ are symmetries of the action. If they are preserved by the ground state they imply $g^{(2)}_{00}(\tilde y)=
g^{(2)}_{11}(\tilde y)$. Now $g^{(2)}_{01}$ is even under the diagonal reflections, while the difference $d=\frac12(\tilde g^{(2)}_{00}-\tilde g^{(2)}_{11})$ is odd. 

Rotations by $\pi/2$ can be obtained by combining reflections. For example, the combination $T_0\tilde D_+$ amounts to a rotation around the origin with an angle $-\pi/2$,
\be\label{S20}
TD_+\tilde y=(\tilde y^1,-\tilde y^0).
\ee
With respect to the $\pi/2$-rotations both $d$ and $g_{01}$ are odd, while $h=\tilde g_{00}+\tilde g_{11}$ is even. 

Symmetries of the action that do not leave the stripe configuration \eqref{S1}, \eqref{S2} invariant lead to equivalent stripes. Equivalent stripes differ from the stripe \eqref{S1}, \eqref{S2} in position, orientation and phases, while they have the same value of the action. (If the ground state is given by one particular stripe configuration the symmetry transformations leading to equivalent, but not identical, stripes are spontaneously broken.) As an example, the action of the stripe is not modified if $\varphi_1$ and $\varphi_2$ are multiplied by global phases $e^{i\beta_1}$ and $e^{i\beta_2}$, respectively. Similarly, the stripes can be displaced by an arbitrary number of units $E_0$ or $E_1$, or they can be rotated by $\pi/2$. Equivalent stripes also obtain from reflections $P,T,D_\pm$ if those are accompanied by $\varphi_1\leftrightarrow\varphi_2$ or $\varphi_1\leftrightarrow\varphi^*_1$. Pure coordinate reflections of stripes do not lead to equivalent stripes, however. For a pure coordinate reflection of the stripe \eqref{S1}, \eqref{S2} the action becomes positive ${\cal L}\y=1/(2\Delta^2)$ - such reflected stripes correspond to a maximum rather than a minimum of the action. Among the symmetries of the action that lead to equivalent (but not identical ) stripes is the flavor rotation
\be\label{35A}
R_f:\quad \varphi_1\to \varphi_2~,~\varphi_2\to\varphi^*_1,
\ee
and the charge conjugation:
\be\label{35A2}
R_c:\quad \varphi_1\to \varphi^*_1~,~\varphi_2\to\varphi^*_2.
\ee
They define equivalence classes of stripes that cannot be rotated into 
each other by phase rotations. While $R_c$ maps $ s_i^+ \leftrightarrow 
s_i^-$ in eq. (\ref{opdef}), the maps 
between $s_1^\pm $ and $s_2^\pm$ can be achieved by $ R_f$.

For $\beta\to\infty$ the ground state is indeed given by a particular
stripe and the symmetries of flavor rotations are spontaneous
broken. This issue is more delicate for finite values of $\beta$. Now
the Mermin-Wagner theorem\cite{MW} forbids any spontaneous breaking of
a continuous symmetry in two dimensions. Nevertheless, for all
practical purposes the system behaves for large enough $\beta$ as if
the flavor symmetries were spontaneously broken. This issue is similar
to the Kosterlitz-Thouless\cite{KT} transition where in the low temperature
phase a mode with all the properties of a Goldstone boson exists. 
For the Kosterlitz-Thouless transition the 
general aspects are well understood by applying the functional
renormalization group to linear and non-linear $\sigma$-models in two
dimensions \cite{FRGKT}. The Mermin-Wagner theorem looses its practical
applicability. While it remains formally valid in the infinite volume
limit, the effects of spontaneous symmetry breaking occur for a
macroscopic system with arbitrary (but finite) size. 
We expect an analogous situation for our model, with 
additional complexity due to the lack of rotation or Lorentz-symmetry 
in the continuum limit.


The limit $\beta\to\infty$ projects on the ``ground state'' for which the action takes its minimum value. For our model the ground state can be solved exactly - it is the stripe configuration \eqref{S1}, \eqref{S2} or an equivalent stripe. In order to show this we first note that any configuration for which ${\cal L}\y$ in eq. \eqref{S6} takes its minimum value for every cell must be a minimum of the action. We next show that this is the case for the stripe. 

Using in eq. \eqref{S7} the definitions \eqref{I3}, \eqref{I4}, the cell action $2\Delta^2{\cal L}\y$ can be written as a sum of terms $\varphi^*_\alpha\varphi_\beta$ of the eight complex fields $\varphi_\alpha$ in the cell. We want to show $2\Delta^2{\cal L}\y\geq -1$. On everyone of the four points in the cell the condition \eqref{I1} must be obeyed, leaving us with $12$ angles. The flavor symmetry ensures that ${\cal L}\y$ does not depend on the overall phases of $\varphi_1$ and $\varphi_2$, such that $10$ angles remain. Finding the minimum of a function of ten angles by analytical means is rather involved. We therefore have proceeded to a numerical evaluation of $2\Delta^2{\cal L}\y$ for random choices of the angles. For $10^{10}$ configurations we have found no value smaller than $-1$ within the numerical accuracy. Together with the value (\ref{S9}) for the 
stripe configuration 
we consider this as sufficient evidence that the stripes \eqref{S1}, \eqref{S2} minimize the action.

\section{Zweibein}
\label{Zweibein}

It is possible to express the action \eqref{I2} in terms of a ``zweibein'' $e_\mu{^m}$. We first introduce
\ba\label{ZA}
\tilde E_\mu{^0}&=&\sqrt{2}i\Delta(\bar\varphi^*_1\partial_\mu\varphi_1-\bar\varphi_1\partial_\mu\varphi^*_1),\nn\\
\tilde E_\mu{^1}&=&\sqrt{2}i\Delta(\bar\varphi^*_2\partial_\mu\varphi_2-\bar\varphi_2\partial_\mu\varphi^*_2),
\ea
such that 
\ba\label{ZB}
S&=&-\frac{\beta}{2\Delta^2}\int d^2x\epsilon^{\mu\nu}\tilde E_\mu{^0}\tilde E_\nu{^1}\nn\\
&=&-\frac{\beta}{4\Delta^2}\int d^2 x\epsilon^{\mu\nu}\epsilon_{mn}\tilde E_\mu{^m}\tilde E_\nu{^n}\nn\\
&=&-\frac{\beta}{2\Delta^2}\int d^2 x\tilde E,
\ea
with 
\be\label{ZC}
\tilde E=\det (\tilde E_\mu{^m}).
\ee
In the continuum limit $\tilde E_\mu{^m}$ transforms as a covariant vector. The form \eqref{ZB} makes the diffeomorphism symmetry of the continuum theory particularly apparent. 

It is instructive to evaluate $\tilde E_\mu{^m}$ for the stripe configuration \eqref{S1}, \eqref{S2}. One obtains from eq. \eqref{S8}
\be\label{ZD}
\tilde E_\mu{^m}=\frac{1}{\sqrt{2}}\left
(\begin{array}{rrr}1&,&-1\\1&,&1\end{array}\right).
\ee
(With $\tilde E=1$ this reproduces ${\cal L}\y=-1/(2\Delta^2)$ in eq. \eqref{S9}.) We may bring the zweibein to a diagonal form by a suitable rotation 
\be\label{ZE}
\tilde e_\mu{^m}=\tilde E_\mu{^n}R_n{^m}.
\ee
For $\tilde E_\mu{^m}$ of the form \eqref{ZD} one needs
\be\label{ZF}
R=\frac{1}{\sqrt{2}}
\left(\begin{array}{rrr}1&,&1\\-1&,&1\end{array}\right)~,~R^TR=1,
\ee
such that the stripe configuration yields $\tilde e_\mu{^m}=\delta^m_\mu$.
We will use this convention and define 
\ba\label{ZG}
\tilde e_\mu{^0}&=&\frac{1}{\sqrt{2}}(\tilde E_\mu{^0}-\tilde E_\mu{^1})=2\Delta Re(i\varphi^*_1\partial_\mu\varphi_1)-
(\varphi_1\leftrightarrow\varphi_2)\nn\\
\tilde e_\mu{^1}&=&\frac{1}{\sqrt{2}}(\tilde E_\mu{^0}+\tilde E_\mu{^1})=2\Delta Re(i\varphi^*_1\partial_\mu\varphi_1)+
(\varphi_1\leftrightarrow\varphi_2).\nn\\
\ea
The action retains the form of a determinant of the zweibein
\be\label{42A}
S=-\frac{\beta}{2\Delta^2}\int d^2 x~\tilde e~,~\tilde e=\det (\tilde e^m_\mu).
\ee

A nonvanishing expectation value of the zweibein indicates the spontaneous breaking of symmetries of the action. This is closely related to spontaneous symmetry breaking by a given stripe configuration. We can use such symmetries in order to transform among equivalent zweibeins. For example, the flavor rotation \eqref{35A} induces
\be\label{42B}
R_f:\quad \tilde e_\mu{^0}\to\tilde e_\mu{^1}~,~\tilde e_\mu{^1}\to-\tilde e_\mu{^0}.
\ee
(This amounts to a $\pi/2$-rotation among the flavor indices of the zweibein. Comparing with eq. \eqref{ZF} we note $R_f=-R^2$.) Similarly, charge conjugation 
\eqref{35A2} induces
\be\label{42B2}
R_c:\quad \tilde e_\mu{^m} \to -\tilde e_\mu{^m},
\ee
such that zweibeins with opposite sign are equivalent.

As usual, the zweibein can be used to define a metric
\ba\label{ZDa}
\tilde g^{(S)}_{\mu\nu}=\hat\eta_{mn}\tilde e_\mu{^m}\tilde e_\nu{^n}
\ea
with $\hat\eta_{mn}=\hat\eta_{nm}$. We define the ``euclidean metric'' $\tilde g^{(E)}_{\mu\nu}$ by choosing $\hat\eta_{mn}=\delta_{mn}$,
\ba\label{ZEa}
\tilde g^{(E)}_{\mu\nu}=\tilde e_\mu{^0}\tilde e_\nu{^0}+\tilde e_\mu{^1}\tilde e_\nu{^1}.
\ea
This coincides with eq. \eqref{9D}. For the stripe configuration
\eqref{S1}, \eqref{S2} one finds $\tilde
g^{(E)}_{\mu\nu}=\delta_{\mu\nu}$. Similar to $\tilde
g^{(2)}_{\mu\nu}$ in eq. \eqref{I8} this metric observable is
independent of the flavor phases, and remains unchanged 
for $\varphi_1\leftrightarrow \varphi_2$ or
$\varphi_i\leftrightarrow\varphi^*_i$. In contrast to eq. \eqref{I8}
it involves four powers of fields $\varphi_i$, however.

Another possibility uses $\hat\eta_{mn}=\eta_{mn}=diag(-1,1)$. The metric corresponding to eq. \eqref{9AA},
\ba\label{GMdef}
\tilde g^{(M)}_{\mu\nu}&=&-\tilde e_\mu{^0}\tilde e_\nu{^0}+\tilde e_\mu{^1}\tilde e_\nu{^1},
\ea
takes for the stripe configuration the values $\tilde g^{(M)}_{\mu\nu}=\eta_{\mu\nu}$. For large enough $\beta$ one therefore expects that the expectation value $\kl g^{(M)}_{\mu\nu}\kr$ has a Minkowski signature, $\kl g^{(M)}_{\mu\nu}\kr=N^{(M)}(\beta)\eta_{\mu\nu}$. We observe that $g^{(M)}_{\mu\nu}$ is again invariant with respect to phase changes of $\varphi_1$ or $\varphi_2$ and to the interchange $\varphi_1\leftrightarrow\varphi_2$. It changes sign, however, under the exchange $\varphi_1\leftrightarrow\varphi^*_1$. 
In consequence, the diagonal reflections $\tilde D_\pm$ change
$g^{(M)}_{00}\leftrightarrow -g^{(M)}_{11}$ while $g^{(M)}_{01}$ is
odd. Thus
$\eta^{\mu\nu}g^{(M)}_{\mu\nu}=-g^{(M)}_{00}+g^{(M)}_{11}$ is
even and $g^{(M)}_{00}+g^{(M)}_{11}$ odd. This additional minus sign
extends to $\pi/2$-rotations under which $-g^{(M)}_{00}+g^{(M)}_{11} $ 
and $g^{(M)}_{01}$
are even and $g^{(M)}_{00}+g^{(M)}_{11}$ is odd. The expectation values 
of the metric (\ref{9AA}) or (\ref{GMdef}) singles out a time direction.
This direction depends on the stripe orientation. 
The transformation (\ref{42B}) switches between time and space directions.
Thus the difference between space and time occurs as an effect of spontaneous
symmetry breaking.

\begin{figure}[htb]
\begin{center}
\includegraphics[width=0.45\textwidth]{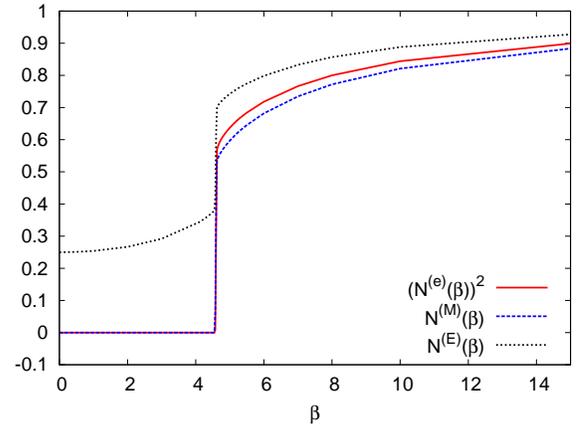}
\caption{ Expectation values of the zweibein and metrics, 
as expressed by the proprtionality factors
$(N^{(e)})^2$, $N^{(M)}$ and $N^{(E)}$, as a function of $\beta$.}
\label{metric_disconnected}
\end{center}
\end{figure}

We have evaluated the expectation value of the zweibein and 
find in the striped phase 
for $\beta>\beta_c$, 
\be\label{K1}
e_\mu\m=\kl\tilde e_\mu\m\kr=N^{(e)}(\beta)\delta^m_\mu.
\ee
This holds if the striped phase is charaterized by the configuration 
(\ref{S1}), (\ref{S2}).
The expectation values in the equivalent striped phases are given by
transformations similar to eqs. (\ref{42B}) and \eqref{42B2}.  
As usual, a metric can be defined as a quadratic expression in the zweibein
\be\label{K2}
g^{(Me)}_{\mu\nu}=e_\mu\m e_\nu{^n}\eta_{mn}=
\big((N^{(e)}(\beta)\big)^2\eta_{\mu\nu}.
\ee
It differs from the Minkowski metric $\kl \tilde g^{(M)}_{\mu\nu} \kr$, cf. 
 eq. (\ref{GMdef}), by the contribution of local zweibein fluctuations. In Fig.~\ref{metric_disconnected} we display $\big(N^{(e)}(\beta)\big)^2$ together with $N^{(M)}(\beta)$. The difference is small and we can approximate $g^{(M)}_{\mu\nu}$ by $g^{(Me)}_{\mu\nu}$ with good accuracy. 

We can go one step further, and try to see if the zweibein expectation
value is well described by the disconnected part of the scalar correlator, 
that is
\ba
{e^{(\varphi)}}_\mu{^0}&=& 2\Delta Re(i\langle\varphi^*_1\rangle\langle\partial_\mu\varphi_1\rangle)-
(\varphi_1\leftrightarrow\varphi_2),\nn\\
{e^{(\varphi)}}_\mu{^1}&=&2\Delta Re(i\langle\varphi^*_1\rangle\langle\partial_\mu\varphi_1\rangle)+
(\varphi_1\leftrightarrow\varphi_2).\nn\\
\ea
Here $\langle \varphi_i \rangle \equiv \langle \bar \varphi(\tilde y ) 
\rangle $ denote the averages of the cell averaged scalar fields,
which are space dependent in the broken phase,
 and $ \langle \partial_\mu \varphi_i ( \tilde y ) \rangle $ are 
averages of lattice derivatives.

\begin{figure}[htb]
\begin{center}
\includegraphics[width=0.45\textwidth]{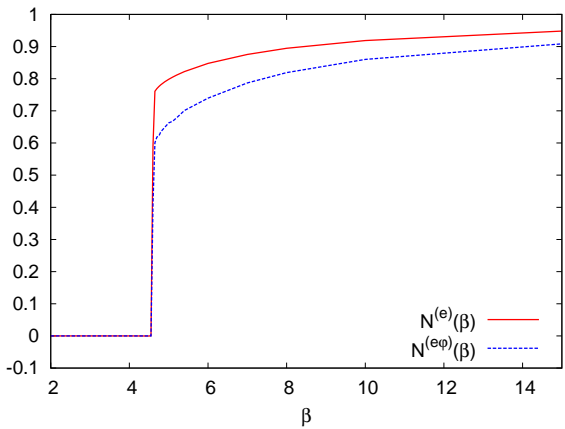}
\caption{Proportionailty factors $N^{(e)}(\beta)$ and 
$  N^{(e\varphi)} ( \beta ) $, 
defined in eqs. (\ref{K1}) and (\ref{nefdef}), as a function of $\beta$. 
 }
\label{bein_disconnected}
\end{center}
\end{figure}

This quantity has the same structure as $ e^m_\mu$,
\be \label{nefdef}
{e^{(\varphi)}}_\mu^m = N^{(e\varphi)} ( \beta ) \delta ^m _\mu. 
\ee
In Fig.~\ref{bein_disconnected} we compare $ N^{(e)}(\beta)$ to 
$  N^{(e\varphi)} ( \beta ) $, and note that the disconnected 
zweibein describes the full zweibein to a reasonable approximation, and the 
quality of the agreement gets better at larger $\beta$.

In the continuum limit the presence of order parameters $e_\mu\m$
singles out particular directions since both $e_\mu{^0}$ and
$e_\mu{^1}$ transform as a vector. A rotation of the axes away from
the $x^0$- and $x^1$-direction does not leave the zweibein diagonal,
such that the diagonal form \eqref{K1} singles out a particular
coordinate frame. The same holds for Lorentz-transformations. In the
usual vielbein formulation of geometry \cite{CAR} the lack of rotation - or
Lorentz- symmetry for any given value of the vielbein is compensated
by an accompanying transformation of the Lorentz frame. A simultaneous
rotation between the vectors $e_\mu{^0}$ and $e_\mu{^1}$, together
with a coordinate rotation, leaves the ``ground state vielbein'' of
the type \eqref{K1} invariant. This guarantees rotation symmetry for
flat euclidean space. A similar property with respect to Lorentz
symmetry ensures Lorentz symmetry for Minkowski space.

In the Cartan-formulation the euclidean metric $g_{\mu\nu}=e_\mu\m
e_\nu{^n}\delta_{mn}$ is obtained from the vielbein by contraction with
the invariant tensor $\delta_{mn}$, such that the metric is
rotation invariant. Then the metric correlation functions
transform under rotations covariantly as dictated by their tensor
properties, provided that there is no other source of rotation
symmetry breaking. Similar statements hold for the Lorentz covariance
of the metric $g^{(Me)}_{\mu\nu}$ \eqref{K2} and its correlation
functions since $\eta_{mn}$ is a Lorentz invariant.

The invariance of the zweibein under combined space and internal 
rotations is, however, not sufficient in 
order to guarantee a rotation invariant setting. In our model the 
expectation value of scalar fields in the form of stripes breaks 
the rotation symmetry without the possibility of an internal
compensating transformation.
As a result, a violation of rotation symmetry for the metric correlation
functions in the continuum becomes plausible, and we discuss this issue
in sect. \ref{Correlation functions}.

Furthermore, in our setting the existence of appropriate internal rotation or Lorentz
transformations among the zweibein components $e_\mu{^0}$ and $e_\mu{^1}$ is not
guaranteed a priori. We investigate this question in the next
section. There we find that no transformation of the field variables
can account for a euclidean rotation among $e_\mu{^0}$ and
$e_\mu{^1}$. In consequence, we expect for the continuum limit the
presence of ``preferred axes'' and a violation of the continuous
rotation symmetry. The issue of Lorentz symmetry is more subtle and
will be discussed in the next section.

\section{Generalized Lorentz transformations}
\label{Generalized Lorentz transformations}

The formulation of the action \eqref{42A} in terms of the zweibein suggests an investigation if a continuous Lorentz type symmetry could be present. Such a symmetry would have to act among the different components of the scalar field. Indeed, if we can find a transformation of the scalars such that $(\tilde e_\mu{^0},\tilde e_\mu{^1})$ transform as a two-component vector (with index $m$), then the contraction with $\epsilon_{mn}$ in eq. \eqref{ZB} yields an invariant. This holds for ``generalized Lorentz-transformations'' corresponding to the groups $SO(2)$ or $SO(1,1)$. (We employ here the name of ``generalized Lorentz transformation'' since they act on the zweibein in analogy to the usual Lorentz transformations. No fermions or spinor representations are involved, however, in our setting.)

Expressing the two complex scalars $\varphi_i$ in terms of four real components $\psi_\alpha$, 

\ba\label{L1}
\varphi_1=\psi_1+i\psi_3~,~\varphi_2=\psi_2+i\psi_4~,~\sum^4_{\alpha=1}\psi^2_\alpha=1,
\ea
we can write the zweibein as
\be\label{L2}
\tilde e_\mu\m=2\Delta\psi_\alpha(\sigma^m)_{\alpha\beta}\partial_\mu\psi_\beta,
\ee
with real antisymmetric $4\times 4$ matrices
\be\label{L3}
\sigma^0=
\left(\begin{array}{rrrr}0&0&-1&0\\0&0&0&1\\1&0&0&0\\0&-1&0&0\end{array}\right)
~,~\sigma^1=
\left(\begin{array}{rrrr}0&0&-1&0\\0&0&0&-1\\1&0&0&0\\0&1&0&0\end{array}\right).
\ee
Under infinitesimal transformations
\be\label{L3A}
\delta\psi=\alpha A\psi
\ee
the zweibein transforms as
\be\label{L4}
\delta e_\mu\m=2\Delta\alpha\psi(\sigma^m A+A^T\sigma^m)\partial_\mu\psi.
\ee
A realization of the Lorentz group $SO(1,1)$ requires 
\be\label{L5}
\sigma^0A+A^T\sigma^0=\sigma^1~,~\sigma^1 A+A^T\sigma^1=\sigma^0,
\ee
such that 
\be\label{L6}
\delta \tilde e_\mu{^0}=\alpha \tilde e_\mu{^1}~,~\delta \tilde e_\mu{^1}=\alpha \tilde e_\mu{^0}.
\ee
The most general solution of eq. \eqref{L5} is 
\be\label{L7}
A=
\left(\begin{array}{cc}
\beta+(1-\gamma)\tau_3~,~\epsilon+\eta{\tau_3}\\
\zeta+\chi{\tau_3}~,~-\beta+\gamma{\tau_3}
\end{array}\right),
\ee
with arbitrary coefficients $\beta,\gamma,\epsilon,\eta,\zeta,\chi$ and Pauli matrices $\tau_k$. On the other hand, a realization of $SO(2)$ would replace the first equation \eqref{L5} by $\sigma^0A+A^T\sigma^0=-\sigma^1$, while keeping the second one unchanged. No solution for $A$ exists in this case. 
At this stage an internal Lorentz transformation remains possible, while 
internal euclidean rotations cannot be realized for our setting of scalar 
fields.

For no choice of parameters $\beta\dots \chi$ the matrix $A$ in eq. \eqref{L7} is antisymmetric. The infinitesimal transformation \eqref{L3A} does therefore not respect the condition $\psi_\alpha\psi_\alpha=1$ for all allowed values $\psi_\alpha$. It is not a genuine symmetry of our model. The basic reason is that $SO(1,1)$ is a noncompact group, while any constraint on the $\psi_\alpha$ that leads to a compact manifold (as the sphere $S^3$ in our case) admits no noncompact isometries. The issue may be understood in more detail by specializing to $A=\frac12 diag(1,-1,1,-1)$ or $\gamma=1/2,\beta=\epsilon=\eta=\zeta=\chi=0$. The transformation \eqref{L3A} induces a change in the length of the vector $\psi_\alpha$, 
\be\label{L8}
\delta(\psi_\alpha\psi_\alpha)=\alpha(\psi^2_1-\psi^2_2+\psi^2_3-\psi^2_4).
\ee
This does not vanish for arbitrary configurations which obey $\psi_\alpha\psi_\alpha=1$. Nevertheless, it vanishes for the subclass of configurations which obey $\psi^2_1+\psi^2_3=\psi^2_2+\psi^2_4$ or $|\varphi_1|^2=|\varphi_2|^2$. For example, this condition is obeyed for the stripe configuration \eqref{S1}, \eqref{S2}. 

This observation has an interesting consequence. For any stripe configuration \eqref{S1}, \eqref{S2} we can infinitesimally increase all $|\varphi_1|$ and decrease all $|\varphi_2|$, such that $|\varphi_1|^2+|\varphi_2|^2$ remains unity for all lattice points. Keeping the phases \eqref{S2} fixed this does not change the action. Such a transformation amounts to an infinitesimal Lorentz rotation among the zweibein components $e_\mu\m$ according to eq. \eqref{L6}.

We summarize that the action \eqref{I2} is invariant under Lorentz
transformations of the group $SO(1,1)$ \be\label{L1a}
\delta{\varphi_1}(x)=\frac{1}{2}\alpha (x)\varphi_1(x)~,~\delta
\varphi_2(x)=-\frac12\alpha(x)\varphi_2(x).  \ee For the continuum
action this symmetry is even a local symmetry, with transformation
parameter $\alpha(x)$ depending arbitrarily on the position
$x$. Indeed, the inhomogeneous part $\sim \partial_\mu\alpha$
resulting from the derivatives cancels. The local character of the
Lorentz symmetry is, however, not respected by the lattice regularization - the
lattice action is invariant only for constant $\alpha$. Furthermore,
the non-linear constraint \eqref{I1} is not compatible with the
Lorentz symmetry.

\section{Metric Correlation functions and symmetries}
\label{Correlation functions}

In this section we investigate the behaviour of fluctuations 
of the metric around the expectation value,  
\be h_{\mu\nu}(x)=\tilde g_{\mu\nu}(x)-g_{\mu\nu}.
\ee
The correlation function, \be\label{I10}
G_{\mu\nu,\rho\sigma}(x,y)=\kl h_{\mu\nu}(x)h_{\rho\sigma}(y)\kr, \ee
characterizes the response of the metric to a source (energy momentum
tensor) in linear order. In our model we find in the metric sector a
non-trivial scaling behavior without a finite correlation length.
\begin{figure}[htb]
\begin{center}
\includegraphics[width=0.44\textwidth]{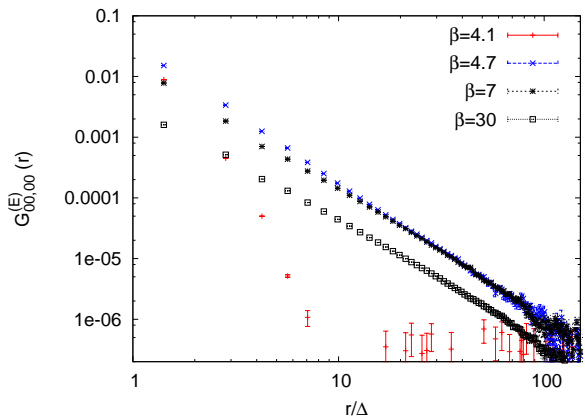}
\caption{Euclidean metric correlator $G^{(E)}_{00,00}$ for different values of $\beta$, parallel to the $E_0$ axis.}
\label{qgtd:fig6}
\end{center}
\end{figure}
\begin{figure}[htb]
\begin{center}
\includegraphics[width=0.44\textwidth]{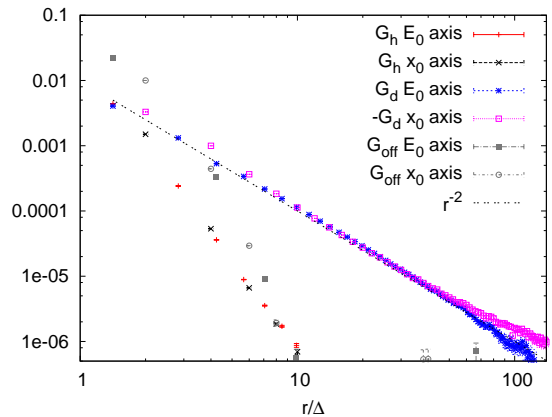}
\caption{Diagonal euclidean metric correlators, 
defined in eq. (\ref{I11}), at $\beta=10$.}
\label{qgtd:fig6_hd}
\end{center}
\end{figure}
In Fig. \ref{qgtd:fig6} we exhibit 
the two point function $ G_{00,00}$ for the euclidean metric (\ref{9D}).
We observe a power law decay 
with the exponent $ \alpha=2 $ at larger couplings, while close to 
$\beta_c$ the power is slightly bigger. 

To disentangle this behavior 
we must decompose this correlator with respect to the discrete symmetries 
of the system. We list in Table \ref{tab:table1} the behavior 
of the various metric 
components with respect to the discrete lattice symmetries, with $+$ for even and $-$ for odd. Here the eigenstates of the reflection symmetries are denoted by $g_{01}$, $h=h_{00}+h_{11}$ and $d=\tilde h_{00}=h_{00}-\frac12 h\delta_{00}=\frac12(h_{00}-h_{11})$. Correlation functions involving two powers of a given eigenstate are even with respect to all discrete symmetries. As result, the correlations
\begin{table}
\begin{center}
\begin{tabular} 
{|c|c|c|c|}\hline
&&&\\
$(E)$&$\frac{\pi}{2}$-rotations&$P,T$&$D_\pm$\\
&&& \\\hline
$h=h_{00}+h_{11}$&$+$&$+$&$+$\\
$d=\frac12(h_{00}-h_{11})$&$-$&$+$&$-$\\
$h_{01}$&$-$&$-$&$+$\\ \hline
\end{tabular}
\end{center}
\caption{Symmetry properties of the euclidean metric with respect to
the discrete symmetries of the lattice.}
\label{tab:table1}
\end{table}

\ba\label{I11}
G_h(x-y)&=&\kl h(x)h(y)\kr\nn\\
&=&G_{00,00}+G_{00,11}+G_{11,00}+G_{11,11},\nn\\
G_d(x-y)&=&\kl d(x)d(y)\kr\nn\\
&=&\frac14(G_{00,00}-G_{00,11}-G_{11,00}+G_{11,11}),\nn\\
G_{off}(x-y)&=&\kl h_{01}(x)h_{01}(y)\kr=G_{01,01},
\ea
are invariant if the difference $y-x$ is rotated by an angle $\pi/2$. 
We plot these correlation functions a function of distance 
in Fig.~\ref{qgtd:fig6_hd}. We observe that the decay of $G_d$ 
for large $(x-y)$ 
is reasonably fitted by a power law
\be\label{I12}
G_d(x-y)=A_d|x-y|^{-\alpha}
\ee
with $\alpha=2$, while $G_h$ and $G_{off}$ show a short distance behaviour with 
a steeper decay. The non-diagonal correlation functions of the euclidean metric 
described below show 
also a short distance behaviour 
The correlator $ G_{00,00}(x,y) $ is thus a linear combination
of the correlators above, some of which have short distance behaviour, 
and some has a long distance behaviour. For $\beta$  close to $\beta_c$
the short distance contributions change the short distance behaviour 
slightly, while for larger $\beta$ the correlator is dominated by 
$ G_d $, which has a decay exponent independent of $\beta$.
 
In the continuum limit for $ r \gg \Delta$ the correlator $ G_d$ 
is the only non-vanishing correlation function
on the $x_0$ and $E_0$ axes for the euclidean metric $ g^{(E)}_{\mu\nu}$.
The tensor properties of the metric imply 
that for a $ \pi/4$-rotation $G_d$ should be 
replaced by $G_{off}$ if rotation symmetry is realized. This is clearly 
not the case, therefore the euclidean
rotation symmetry is broken in the continuum limit.
The behavior of the correlation functions is related to an expansion
of  
the effective action in second order in the fields . 
Our findings clearly indicate that the effective action is not 
of the simple form discussed in the Appendix \ref{appa}.

Using the discrete symmetries above one can find that certain 
correlation functions have to vanish, as described below.
For example, the correlation function 
\be\label{32A}
\kl d(x) h_{01}(y)\kr=\frac12 (G_{0001}-G_{1101}).
\ee
is even under $\pi/2$-rotations.
Since this correlation is odd under $P,T$ and $D_\pm$ it has to vanish on the $x^0$- and $x^1$-axes, as well as on the diagonal axes $\sim E_0$ or $\sim E_1$. 
On the other hand, the correlations
\be\label{32B}
G_{hd}(x-y)=\kl h(x)d(y)\kr = G_{00,00}-G_{11,11}
\ee
and
\be\label{32C}
G_{hoff}(x-y)=\kl h(x)g_{01}(y)\kr =
G_{00,01}+G_{11,01}
\ee
change sign if $y-x$ is rotated by $\pi/2$. The correlation $G_{hd}$ vanishes if $y-x$ is parallel to one of the diagonal axes $\sim E_0$ or $\sim E_1$, while $G_{hoff}$ is zero if $y-x$ is parallel to the $x^0$- or $x^1$-axis. 
These features are confirmed by our numerical results.

Equivalently, we can investigate the action of the discrete symmetries
directly for the correlations (\ref{I10}). For example, the
correlation functions of the type
\be\label{S14}
G_{\mu\nu01}(\tilde y)=\kl h_{\mu\nu}(0)h_{01}(\tilde y)\kr
\ee
have to obey
\ba\label{S15}
G_{0001}(T\tilde y)&=&G_{0001}(P\tilde y)=-G_{0001}(\tilde y),\nn\\
G_{1101}(T\tilde y)&=&G_{1101}(P\tilde y)=-G_{1101}(\tilde y),\nn\\
G_{0101}(T\tilde y)&=&G_{0101}(P\tilde y)=G_{0101}(\tilde y),
\ea
for 
\be\label{S16}
\tilde y=(\tilde y^0,\tilde y^1)~,~T\tilde y=(-\tilde y^0,\tilde y^1)~,~P\tilde y=(\tilde y^0,-\tilde y_1).
\ee
In particular, the correlations $G_{0001}$ and $G_{1101}$ have to vanish on the $x^0$- and $x^1$-axes, i.e. for $\tilde y=(\tilde y^0,0)$ or $\tilde y=(0,\tilde y^1)$. 
For $\tilde y$ on one of the diagonal axes, e.g. for $\tilde y=(m,m)$ or $\tilde y=(m,-m)$ this implies for the correlations
\ba\label{S19}
\kl h_{00}(0) d(\tilde y)\kr&=&-\kl h_{11}(0)d\y\kr,\nn\\
\kl h_{01}(0)d\y\kr&=&0.
\ea
The symmetry arguments are the same for all euclidean metrics -
they hold both for $ h^{(2)}_{\mu\nu}$ and $ h^{(E)}_{\mu\nu}$.

\begin{figure}[htb]
\begin{center}
\includegraphics[width=0.44\textwidth]{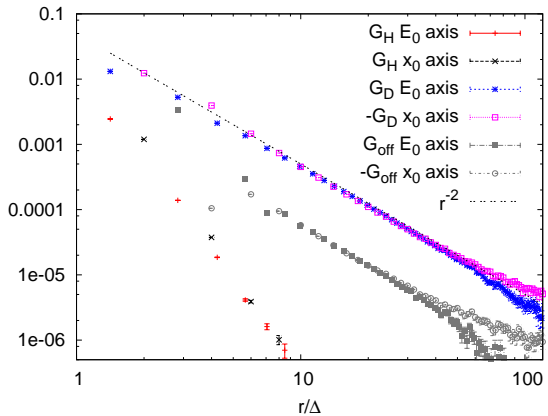}
\caption{Diagonal Minkowski metric correlators,
defined in eq. (\ref{I11_m}), for $\beta=10$.}
\label{qgtd:fig6m_hd}
\end{center}
\end{figure}

The correlation functions for the Minkowski metric behave similarly 
to the euclidean correlations. The decomposition is now done whith 
respect to a Mikowski signature, as displayed in Table \ref{table2}.
\begin{table}
\begin{center}
\begin{tabular}{|c|c|c|c|}
\hline
&&&\\
$(M)$&$\frac\pi2$-rotations&$P,T$&$D_\pm$\\ 
&&&\\ \hline
$H_M=-h^{(M)}_{00}+h^{(M)}_{11}$&$+$&$+$&$+$\\ 
$D_M=-\frac12(h^{(M)}_{00}+h^{(M)}_{11})$&$-$&$+$&$-$\\ 
$h^{(M)}_{01}$&$+$&$-$&$-$\\ \hline
\end{tabular}
\end{center}
\caption{Symmetry properties of the Minkowski metric with respect to 
the discrete symmetries of the lattice.}
\label{table2}
\end{table}
In Fig.~\ref{qgtd:fig6m_hd} we show the diagonal correlations
\ba\label{I11_m}
G_H(x-y)&=&\kl H_M(x)H_M(y)\kr\nn\\
&=&G^{(M)}_{00,00}-G^{(M)}_{00,11}-G^{(M)}_{11,00}+G^{(M)}_{11,11},\nn\\
G_D(x-y)&=&\kl D_M(x)D_M(y)\kr\nn\\
&=&\frac14(G^{(M)}_{00,00}+G^{(M)}_{00,11}
+G^{(M)}_{11,00}+G^{(M)}_{11,11}),\nn\\
G_{off}(x-y)&=&\kl h^{(M)}_{01}(x)h^{(M)}_{01}(y)\kr=G^{(M)}_{01,01}.
\ea

\begin{figure}[htb]
\begin{center}
\includegraphics[width=0.44\textwidth]{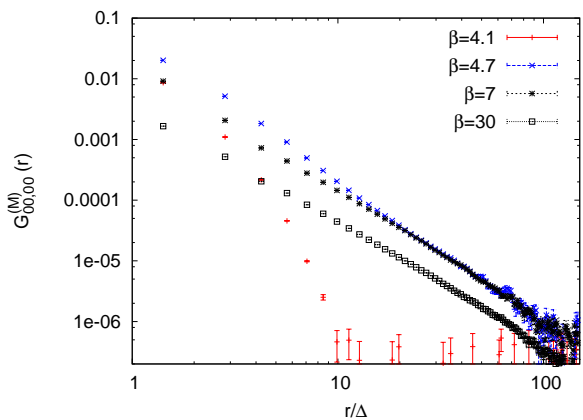}
\caption{Minkowski metric correlator $G^{(M)}_{00,00}$ for different values of $\beta$, parallel to the $E_0$ axis.}
\label{qgtd:fig6m}
\end{center}
\end{figure}

Similarly to 
the euclidean case, $G_D$ shows a long distance behaviour of a power 
law decay with $ \alpha \approx 2$.
Along with $ G_{D,off}$ this is the dominant correlation function 
in the continuum limit.
The correlation function $G_{off}$ for the 
off diagonal metric components
also shows a power law decay and remains present in the continuum limit,
even though its value is smaller the $G_D$ and $G_{D,off}$ by an order of 
magnitude.
On the other hand, $G_H$ decays quickly and plays no role in the 
continuum limit.
In Fig.~\ref{qgtd:fig6m} we show the correlator 
$ G^{(M)}_{00,00}$ which behaves similarly to $G^{(E)}_{00,00} $ shown in 
Fig.~\ref{qgtd:fig6}.

\section{Zweibein correlations}
\label{Zweibein correlations}
The intriguing features of the correlations for the euclidean and Minkowski metric can be understood better in terms of the correlation functions for the zweibein fluctuations
\be\label{Z1}
f_\mu\m=\tilde e_\mu\m-e_\mu\m=\tilde e_\mu\m-
\kl \tilde e_\mu\m\kr.
\ee
As we have discussed above the expectation value $e_\mu\m$ preserves the discrete symmetries $P,T,D_\pm$ and $\pi/2$-rotations. We classify the zweibein fluctuations according to these symmetries
\ba\label{Z2}
s&=&\frac12(f_0{^0}+f_1{^1})~,~a=\frac12(f_0{^0}-f_1{^1}),\nn\\
u&=&\frac12(f_0{^1}+f_1{^0})~,~v=\frac12(f_0{^1}-f_1{^0}).
\ea
The transformation properties of the eigenstates $s,a,u,v$ under $P,T,D_\pm$ and $\pi/2$-rotations are collected in table~\ref{table_zweibein}.

\begin{table}
\begin{center}
\begin{tabular}{|c|c|c|c|}\hline
&&&\\ 
$(Z)$&$\frac\pi2$-rotations&$P,T$&$D_\pm$\\ 
&&&\\ \hline
$s$&$+$&$+$&$+$\\ 
$a$&$-$&$+$&$-$\\ 
$u$&$-$&$-$&$+$\\ 
$v$&$+$&$-$&$-$\\ \hline
\end{tabular}
\end{center}
\caption{ Symmetry properties of zweibein fluctuations.}
\label{table_zweibein}
\end{table}

\medskip
This implies that the correlations $\kl su\kr,\kl sv\kr,\kl au\kr$ and $\kl a v\kr$ must vanish on the $x^0$- and $x^1$- axes, while $\kl sa\kr,\kl sv\kr,\kl au\kr,\kl uv\kr$ are zero on the diagonal axes $\sim E_0$ or $\sim E_1$. A rotation of the axes by $\pi/2$ leads to a minus sign for the correlations $\kl sa\kr, \kl su\kr, \kl av\kr$ and $\kl uv\kr$, while the other combinations are invariant. 

\begin{figure}[htb]
\begin{center}
\includegraphics[width=0.45\textwidth]{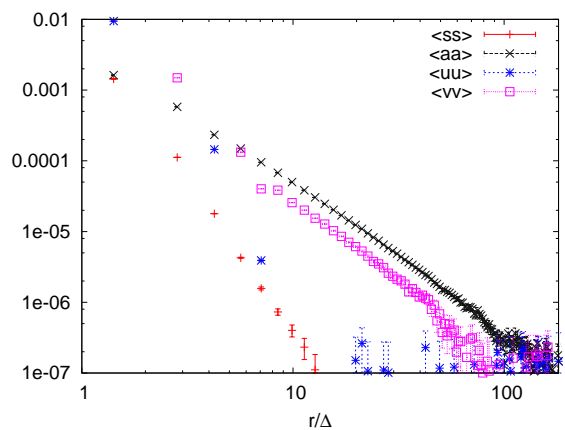}
\caption{Diagonal correlations of zweibein fluctuations on the 
$E_0$ axis measured at $\beta=7$. }
\label{beincorr}
\end{center}
\end{figure}

\begin{figure}[htb]
\begin{center}
\includegraphics[width=0.45\textwidth]{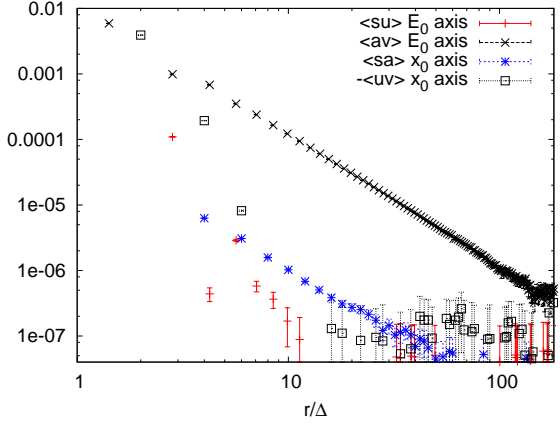}
\caption{Non-vanishing non-diagonal correlations of zweibein fluctuations 
on the $E_0$ and $x_0$ axes measured at $\beta=7$. }
\label{beincorr-nd}
\end{center}
\end{figure}

The zweibein correlators also show a powerlaw decay similarly to the 
metric correlators, as shown in Fig.~\ref{beincorr}. We also plot 
the non-vanishing non-diagonal correlations 
on the $ E_0$ and $x_0$ axes in Fig.~\ref{beincorr-nd}.
Again, only some modes exhibit the slow, power law decay, while other modes 
show 
faster decay, similar to the correlations of the metric tensor.
The correlations relevant for the continuum limit are $ \kl aa \kr $, 
$ \kl vv \kr$ and $\kl av \kr$.


The fluctuations of the metrics \eqref{ZDa} are related to the zweibein fluctuations by 
\ba\label{Z3}
h^{(S)}_{\mu\nu}&=&\tilde e_\mu\m\tilde e_\nu{^n}\hat\eta_{mn}-
\kl \tilde e_\mu\m\tilde e_\nu{^n}\kr\hat\eta_{mn}\nn\\
&=&\{ f_\mu\m f_\nu{^n}-e_\mu\m f_\nu{^n}-
f_\mu\m e_\nu\n
+e_\mu\m e_\nu\n\nn\\
&&-\tilde e_\mu\m\tilde e_\nu\n\}\hat\eta_{mn}.
\ea
Taking the expectation value of eq. \eqref{Z3} and employing $\kl h^{(S)}_{\mu\nu}\kr=0$, $\kl f_\mu\m\kr=0$, relates the difference between the metrics \eqref{9A} and \eqref{K2} to the local zweibein fluctuations. This holds
both for euclidean and Minkowski signature, such that for $ S=M,E$ 
one finds
\be\label{Z4}
g^{(S)}_{\mu\nu}-g^{(Se)}_{\mu\nu}=\kl f_\mu\m f_\nu\n\kr \hat\eta_{mn}.
\ee

Furthermore, if we neglect higher correlations $\sim \kl f^3\kr, \kl f^4\kr_c$ one obtains an approximate relation between the metric correlations for $g^{(S)}_{\mu\nu}$ and the zweibein correlations. (Here $\kl f^4\kr_c \sim \kl \big(f^2(x)-\kl f^2(x)\kr \big) (f^2(y)-\kl f^2(y)\kr\big)\kr$, with indices and $\hat\eta$ omitted.) This leading order relation reads
\ba\label{Z5}
&&\kl h^{(S)}_{\mu\nu}(x) h^{(S)}_{\rho\sigma}(y)\kr \approx
F^{mr}_{\mu\rho} e_{\nu m} e_{\sigma r}\nn\\
&&\qquad +F^{ms}_{\mu\sigma}e_{\nu m} e_{\rho s}+ F^{nr}_{\nu\rho} e_{\mu n} e_{\sigma r}+
F^{ns}_{\nu\sigma} e_{\mu n} e_{\rho s}, 
\ea
where we define 
\be\label{Z6}
F^{mn}_{\mu\nu}=\kl f_\mu\m(x)f_\nu\n(y)\kr
\ee
and 
\be\label{Z7}
e_{\mu n}=e_\mu\m\hat\eta_{mn}=N^{(e)}(\beta)\hat\eta_{\mu n}.
\ee

For the euclidean metric correlation this yields in the 
striped phase given by eq. (\ref{S1})
\ba\label{Z8}
G^{(E)}_{00,00}&=&4 F^{00}_{00}~,~G^{(E)}_{11,11}=4 F^{11}_{11},\nn\\
G^{(E)}_{00,11}&=&G^{(E)}_{11,00}=4 F^{01}_{01}=4 F^{00}_{11},\nn\\
G^{(E)}_{01,01}&=& F^{00}_{11} + 2 F^{01}_{10} + F^{11}_{00}.
\ea
\begin{figure}[htb]
\begin{center}
\includegraphics[width=0.45\textwidth]{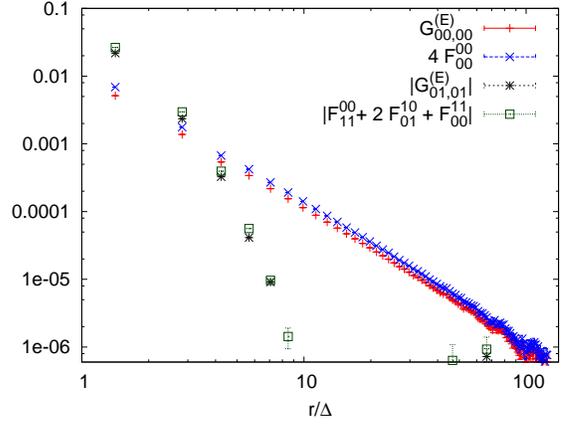}
\caption{Correlations of the euclidean metric compared with estimates 
from zweibein correlations according to eq. (\ref{Z8}) measured at 
$\beta=10$. Since $G^{(E)}_{01,01}$ has an alternating sign,
 we show the absolute value. }
\label{gcorr-bcorr}
\end{center}
\end{figure}
In Fig.~\ref{gcorr-bcorr} we show that the metric correlations are 
well aproximated using the estimate (\ref{Z8}) in terms of the 
 zweibein correlators.
This is true even for the case of a fastly decaying mode.
This evidence means that the higher order zweibein correlations 
quite small. It also implies that the power-law behaviour of the metric
correlations is governed by the power law behaviour of the 
zweibein correlators. 
In Fig.~\ref{beincorr_betas} we show the 
$\langle a a \rangle $ correlator for different values of $\beta$.
Within our precision the decay exponent $ \alpha \approx 2 $ is found 
to be independent of $\beta$ in the stripe phase
for $ \beta > \beta_c $.

\begin{figure}[htb]
\begin{center}
\includegraphics[width=0.45\textwidth]{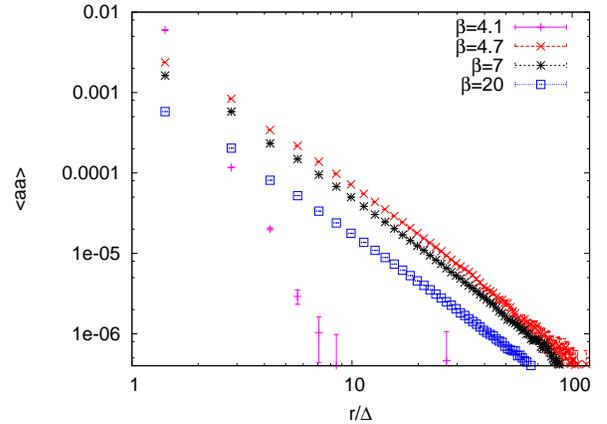}
\caption{Correlations $\kl aa\kr$ of zweibein fluctuations on the 
$E_0$ axis for different values of $\beta$. }
\label{beincorr_betas}
\end{center}
\end{figure}

\begin{figure}[htb]
\begin{center}
\includegraphics[width=0.45\textwidth]{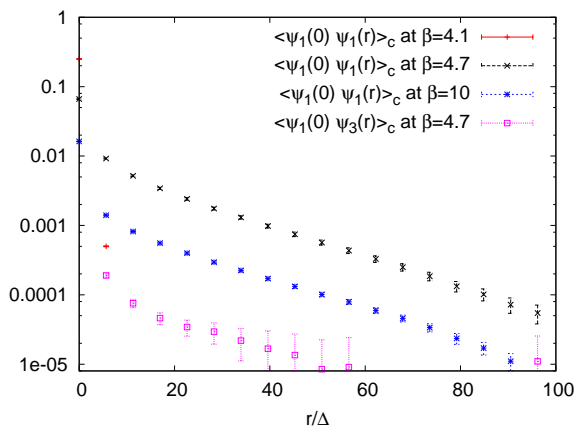}
\caption{Typical scalar correlators along the $E_0$ axis 
measured at $\beta=4.9$. Note that the x axis has a linear scale. }
\label{skcorr}
\end{center}
\end{figure}

Finally, the scalar correlations are qualitatively different from 
the metric and zweibein correlators. Some typical scalar correlators
are shown in Fig.~\ref{skcorr}. As one observes, they 
can be approximated by an exponential decay rather than a 
power law. The decay constant is roughly independent of the coupling above 
$ \beta_c$, and the overall amplitude of the correlations 
decays with increasing $ \beta$.
For the scalar correlator we must take into account 
that the ground state in the broken phase is only invariant 
with respect to translations by $ 4 \sqrt{2} \Delta $.
Therefore we plot
the scalar correlator for distances which are a 
multiple of $ 4 \sqrt{2} \Delta $.

 We have shown in Fig.~\ref{metric_disconnected} that the metric 
expectation value 
is well approximated by disconnected contributions, i.e. the zweibein 
expectation values. In the first part of this section we have shown that 
also
the metric correlators are well described by the zweibein correlators, 
neglecting higher order zweibein correlators. In this sense the 
metric correlations are largely determined by the zweibein.
The relation of the zweibein 
and the scalars seem to be different.
The zweibein expectation value is well described by the 
disconnected part, i.e. the scalar expectation values, as shown 
in Fig.~\ref{bein_disconnected}. 
One may ask if this can be extended to the correlation functions by
neglecting higher order connected correlation functions of the scalars.
We address this issue in appendix 
\ref{Zweibein correlations from scalar correlations}. However, the 
different qualitative behavior of zweibein correlators and scalar 
correlators make it unlikely that 
the zweibein correlator can be approximated 
by a linear combination of the scalar correlators.

\section{Discussion and Conclusions}
\label{Conclusions}

In this paper we have investigated the non-linear $\sigma$-model given
by eqs. (\ref{I1}), (\ref{I2}).
 This model can be interpreted as a
classical statistical model. In this case $\beta$ plays the role of an
inverse temperature. This new class of non-linear $\sigma$-models
shows several interesting features. For all temperatures below the
critical temperature, $\beta>\beta_c$, we find a universal critical
behavior. Our model is therefore an example for self-tuned
criticality, where critical behavior occurs without a tuning of
parameters. This contrasts with the usual situation where critical
behavior occurs only for a special choice of parameters,
e.g. $\beta=\beta_c$. The presence of long range correlations for all
$\beta>\beta_c$ shows certain analogies with the Kosterlitz-Thouless
phase transition \cite{KT}. Our model belongs, however, to a 
new universality class that differs in important aspects from the
universality class characterizing the Kosterlitz-Thouless phase
transition. It is not restricted to an abelian symmetry $SO(2)$ and
shows a very different long distance behavior. It remains to be seen
if some particular condensed matter system realizes this universality
class.

A characteristic feature of our model is the stripe phase, where the
symmetries of discrete reflections and $\pi/2$-rotations on the
lattice are preserved only if they are combined with gauge
transformations acting on the flavor indices of the scalar
field. The stripes single out preferred axes for quantities that are
not invariant under gauge transformations. The presence of a stripe
phase is not mainly a property of the lattice formulation. We show in
appendix \ref{Field equations} that a stripe phase is also present in a continuum model
with effective action similar to eq. (2). Such an effective action is
a candidate for the description of the universality class, while we
have not yet made any detailed comparison of its properties with our
numerical results. Nevertheless, we have established that the effective
action discussed in appendix \ref{Field equations} 
can describe the first order phase
transition from the disordered to the stripe phase.

The most prominent property of our non-linear $\sigma$-model is lattice
diffeomorphism invariance. This implies general coordinate invariance
or diffeomorphism symmetry in the continuum limit. Our main interest
concerns the possible relations to quantum gravity and we therefore
focus on collective order parameters that play the role of a metric
and a zweibein. This realizes geometrical features for our statistical
model, close to the conceptual framework of ref. \cite{CWGS}. We have
computed expectation values as well as correlation functions for the
collective fields using numerical simulations.

Our main findings are mentioned in the introduction and 
we may concentrate here on two aspects: (1) non-vanishing 
expectation values of the
metric for the vacuum state, describing flat space with either
Minkowski or euclidean signature, (2) long range correlations for the
metric fluctuations that decay with $r^{-2}$ as a function of
euclidean distance $r$.

As in usual gravity, the geometry of space or spacetime is deformed by
the presence of matter. This can be seen by introducing an energy
momentum tensor as a source term for the metric fluctuations. The
response of geometry to a local energy momentum tensor is fixed in the
linear approximation by the metric correlation function. One finds for
the deviation from flat space 
\be\label{CZA}
h_{\mu\nu}(x)=\int_yG_{\mu\nu,\rho\sigma}(x,y)t^{\rho\sigma}(y), 
\ee
with $t^{\rho\sigma}$ an appropriate energy momentum density and
$G_{\mu\nu,\rho\sigma}$ the metric correlator defined in eq. (\ref{I10}). In
particular, for $t^{00}(y)=K\delta(y)$ and vanishing other components,
one obtains 
\be\label{CZB} h_{\mu\nu}(x)=K G_{\mu\nu,00}(x,0).  
\ee 
For
this source the metric perturbation $h_{00}$, that may be compared to
the Newtonian potential in four dimensional gravity, decays $\sim
r^{-2}$, with euclidean distance $r^2=x^\mu x^\nu\delta_{\mu\nu}$, as
visible in Fig.~\ref{qgtd:fig6}.

A static localized massive object in a Minkowski setting 
would correspond to a time independent energy momentum tensor 
\be
t^{00}( y^0 , y^1 ) = M \delta ( y^1 ). 
\ee
This results in a static metric that decays inversely proportional
to the spatial distance $|x_1|$,
\be
h_{00} ( x_0, x_1 ) \sim {M \over |x_1 | },
\ee
similar to four dimensional gravity.

Our model can be considered as a model for two-dimensional quantum
gravity in the sense that the effective action for the metric is
invariant under general coordinate transformations and that the metric
correlations are long range. As a perhaps surprising effect the vacuum
state corresponds to flat space even in the presence of quantum
fluctuations. There are, however, also important differences as
compared to Einstein gravity in four dimensions. They are mainly
related to the non-trivial stripe order for $\beta>\beta_c$. The
presence of preferred axes may lead to features that are not encountered
if all order parameters preserve Lorentz symmetry as for standard
four-dimensional gravity. The stripe order is reflected in the
behavior of the correlation functions for the zweibein. In appendix E
we discuss a simple ansatz for the zweibein effective action in the
presence of stripes. It seems to differ substantially from the simple
Lorentz-invariant setting of appendix B. While the effective action
for scalars and zweibein is assumed to be Lorentz invariant, the
Lorentz symmetry may be 
spontaneously broken  the vacuum by the stripe
configuration.

Several important issues remain to be solved 
before a more realistic model for quantum 
gravity can be constructed from a suitable scalar field theory 
on a lattice. The approach using a collective 
vielbein seems quite promising. A diffeomorphism invariant lattice 
action can then easily be formulated by
employing the determinant of the collective 
vielbein. In this case one would like to implement the Lorentz 
transformations acting on the internal or flavor index 
of the vielbein as an exact symmetry.
This is possible along the lines discussed 
in appendix \ref{Field equations}.
One may employ a Lorentz invariant potential $ V( \rho_1 \rho_2 ) $
already for the microscopic lattice action. The non-linear 
constraint (\ref{I1}) can then be replaced by a Lorentz invariant 
constraint, for example by a bound on the Lorentz invariant product 
$\rho_1 \rho_2 \le C$. The second issue concerns the preservation
of a global Lorentz symmetry
for a flat space ground state. One possible solution is a non-zero 
expectation value for the vielbein $ e_\mu^m \sim \delta ^m_\mu $,
while the order parameter for the stripe configuration vanishes.
Finally, an important step is the transition from two to four dimensions.

Several of 
the mentioned problems are absent or 
solved in lattice spinor gravity. However, reliable computations 
are difficult for spinor gravity. For this reason it seems worthwhile
to pursue in parallel the scalar approach to lattice gravity 
which permits relatively cheap numerical simulations.

\begin{appendix}
\section{Propagating metric in two-dimensional gravity}
\label{appa}

In this appendix we demonstrate that two-dimensional gravity can
have propagating metric degrees of freedom.  The issue which degrees
of freedom propagate depends on the form of the quantum effective
action.  We present a simple example for such an action where metric
degrees of freedom are indeed propagating.

In two dimensions the curvature tensor has only one independent component that is related to the curvature scalar $R$ by 
\be\label{A}
R_{0101}=\frac12\det (g_{\mu\nu})R.
\ee
The integral $\int_x\sqrt{g}R$ is a topological invariant and 
does not contribute to the field equations of the metric. If only 
this term and a two-dimensional cosmological
constant are present in the effective action there will be no kinetic term 
for the metric, such that the metric is not a propagating field.

Nevertheless, a diffeomorphism invariant effective action can be constructed as
\be\label{AA}
\Gamma=\int_x\sqrt{g}Rf(-D^2)R,
\ee
where $D^2=D^\mu D_\mu$ and $D_\mu$ is the covariant derivative. This can be generalized by adding terms with higher powers of $R$.  We will concentrate on non-local invariants\cite{CWNL} of the type
\be\label{B}
\Gamma=c_\kappa I_\kappa~,~I_\kappa=\int_x\sqrt{g}R(-D^2)^{-\kappa}R,
\ee
with $\kappa>0$. Our setting remains more general, however.

In linear order of an expansion around flat space, $h_{\mu\nu}=g_{\mu\nu}-\eta_{\mu\nu},h=h_{\mu\nu}\eta^{\mu\nu}$ one finds (for arbitrary dimension $d$)
\be\label{C}
R=\partial_\mu\partial_\nu h^{\mu\nu}-\partial^2 h=-\frac{d-1}{d}\partial^2\zeta,
\ee
with
\be\label{D}
\zeta=h-\frac{d}{d-1}\frac{\partial_\mu\partial_\nu}{\partial^2}\tilde h^{\mu\nu}~,~h_{\mu\nu}=\tilde h_{\mu\nu}+\frac1d h\eta_{\mu\nu}.
\ee
It is easy to check that $\zeta$ is invariant under the inhomogeneous part of the gauge transformations, $\delta_{\rm {inh}}h_{\mu\nu}=\partial_\mu\xi_\nu+\partial_\nu\xi_\mu$. For $d=2$ the invariant \eqref{B} reads in quadratic order 
in $h_{\mu\nu}$
\be\label{E}
I_\kappa=\frac14\int_x\zeta(-\partial^2)^{2-\kappa}\zeta.
\ee
For an effective action $\Gamma=c_\kappa I_\kappa,\kappa>1$, and for euclidean signature $\eta_{\mu\nu}=\delta_{\mu\nu}$, this implies correlation functions decoupling for large $(x-y)$ as 
\ba\label{F}
\kl \zeta(x)\zeta(y) \kr \sim |x-y|^{2-2\kappa},\nn\\
\kl \partial^2\zeta(x)\partial^2\zeta(y)\kr\sim 
|x-y|^{-(2+2\kappa)}.
\ea
For Minkowski signature the field equation for $\zeta$, which is obtained 
by taking a funtional derivative of the 
effective action (\ref{E}), describes a relativistic wave equation 
for a propagating degree of freedom.

A non-local effective action of the type \eqref{B} typically indicates the presence of a massless degree of freedom. For a suitable choice of degrees of freedom, $\Gamma$ can often be written in an equivalent (quasi-)local form. 
For example, an effective action 
\be\label{G}
\Gamma=\int_x\sqrt{g}\left\{-\frac12\chi Z(-D^2)(-D^2)\chi+f\chi R\right\}
\ee
describes a scalar field $\chi$ with non-trivial wave 
function renormalization 
$Z$ of the kinetic term and local coupling to the curvature scalar $R$.
 This implies for the scalar field $\chi$ the field equation
\be\label{H}
\chi=fZ^{-1}(-D^2)(-D^2)^{-1}R.
\ee
Insertion of eq. \eqref{H} into eq. \eqref{G} yields the gravitational effective action
\be\label{I}
\Gamma=\frac{f^2}{2}\int_x\sqrt{g}RZ^{-1}(-D^2)(-D^2)^{-1}R.
\ee
For $Z=-D^2$ we recover the action \eqref{B} with $\kappa=2$. For an expansion around flat space the scalar $\chi$ is directly related to $\zeta$ in eq. \eqref{D}
\be\label{J}
\chi=\frac f2 Z^{-1}(-\partial^2)\zeta.
\ee
The scalar $\chi$ or $\zeta$ is the only propagating field in this type of two-dimensional gravity. Correlation functions for metric components with an overlap with this scalar field, e.g. for $h_{00}$ or $h_{11}$, should show a powerlike decay given by eq. \eqref{F}.

The effective action \eqref{B} is invariant under general coordinate transformations. This is reflected in the linear expansion by the fact that only the particular combination $\zeta$ of metric components contributes to the action. Without diffeomorphism symmetry nothing particular distinguishes $\zeta$ from the other components of $h_{\mu\nu}$. 
The expansion of the effective action (\ref{B})
around a flat space exhibits the symmetry of global rotations or global 
Lorentz transformations, depending on the signature of the metric. 
In this important aspect it differs from the effective 
action for the 
collective metric in our non-linear $\sigma$-model (\ref{I2}).

\section{Possible effective action for  
propagating zweibein in two dimensions}
\label{appb}

In this appendix we briefly discuss an example for a possible form of
an effective action for the zweibein. This serves as an illustration
of some of the effects that may be expected for a more realistic
effective action.  We insist on diffeomorphism symmetry of the
effective action, but we do not require local Lorentz symmetry acting
on the index $m$ of the zweibein $e^m_\mu$.
Similarly, we also do not impose euclidean rotation symmetry.
As a consequence, the covariant derivative contains
no spin connection, and $ D_\rho e^m_\mu$ can differ from zero.
(A discussion of this type of generalized geometry can be found in 
ref.~\cite{CWGG}).

Our example for an effective action involves the determinant of a 
``renormalized zweibein'' 
\be\label{G1}
e^{Rm}_\mu=Z^{m\nu}_{\mu n} e_\nu\n.
\ee
Here the ``wave function renormalization'' $Z^{m\nu}_{\mu n}$ is a function of covariant derivatives $D_\mu$ such that $e^{R m}_\mu$ transforms again as a covariant vector. An example is $(D^2=D_\mu D^\mu)$
\be\label{G2}
Z^{m\nu}_{\mu n}= z_1(-D^2)\delta^\nu_\mu\delta^m_n+
z_2(-D^2) D_\mu D^\nu \delta^m_n,
\ee
with $z_1$ and $z_2$ scalar functions.
(Covariant derivatives involve the Levi-Cevita connection in the usual way, but no spin connection.) For the leading term in the effective action our ansatz reads
\ba\label{G3}
\Gamma&=&-\frac{\beta}{4\Delta^2}\epsilon^{\mu\nu}\epsilon_{mn}\int d^2 x e^{Rm}_\mu e^{Rn}_\nu\nn\\
&=&-\frac{\beta}{4\Delta^2}\int d^2 x A^{\mu\nu}_{mn} e_\mu\m e_\nu\n,
\ea
with 
\be\label{G4}
A^{\mu\nu}_{mn}=Z^{r\mu}_{\rho m} Z^{s\nu}_{\sigma n}\epsilon^{\rho\sigma} \epsilon_{rs}.
\ee
For the example \eqref{G2} one has 
\ba\label{G5}
A^{\mu\nu}_{mn}&=&\{z^2_1\epsilon^{\mu\nu}+z_1z_2(D_\rho D^\mu\epsilon^{\rho\nu}-D_\rho D^\nu\epsilon^{\rho\mu})\nn\\
&&+z^2_2 D_\rho D^\mu D_\sigma D^\nu\epsilon^{\rho\sigma}\}\epsilon_{mn}.
\ea

The effective action \eqref{G3}  is diffeomorphism invariant provided that covariant derivatives of tensors are again tensors, such that $A^{\mu\nu}_{mn}$ transforms as a scalar function multiplied by $\epsilon^{\mu\nu}$. The effective action would be invariant under a global generalized Lorentz transformation if the covariant and contravariant derivatives are singlets with respect to this transformation. This is possible only for one of the groups $SO(2)$ or $SO(1,1)$, but not for both simultaneously. For example, a definition 
\be\label{G6}
D^\mu=e_m{^\mu} e_n{^\nu}\hat\eta^{mn}D_\nu=g^{\mu\nu}D_\nu
\ee
requires the specification of $\hat\eta_{mn}$, with inverse $\hat\eta^{mn}$ obeying $\hat\eta^{mn}\hat\eta_{np}=\delta^m_p$. The contravariant derivative also involves the inverse zweibein $e_m{^\mu}$ which is defined by the relations
\be\label{G7}
e_m{^\mu}e_\mu{^n}=\delta^n_m~,~e_\mu\m e_m{^\nu}=\delta_\mu^\nu. 
\ee

In order to gain some intuition for the implications of the effective action \eqref{G3} we first consider the approximation where $z_1$ is a function of $-\partial^2=-\partial_\mu\partial_\nu\hat\eta^{\mu\nu}$, while $z_2=0$. The quantum field equation reads
\be\label{G9}
\frac{\delta\Gamma}{\delta e^m_\mu}=-\frac{\beta z^2_1}{2\Delta^2}\epsilon^{\mu\nu}\epsilon_{mn} e_\nu\n=0. 
\ee
Flat space with $e_\nu\n\sim\delta^n_\nu$ solves this field equation provided that the Fourier transform $z_1(q)$ vanishes for vanishing momenta $q=(q_0,q_1)$. The second functional derivative reads in momentum space
\be\label{G10}
\frac{\delta^2\Gamma}{\delta e_\mu\m(-q)\delta e_\nu\n(q')}=-\frac{\beta z^2_1}{2\Delta^2}
\epsilon_{mn}\epsilon^{\mu\nu}
\delta(q,q').
\ee
For $z_1$ a function of $q^2=q^\mu q_\mu=- q^2_0+q^2_1$ the corresponding inverse propagator is a Lorentz covariant expression. 

In the space of the four-component vector $E=(e_0{^0},e_0{^1},e_1{^0},e_1{^1})$ the inverse propagator takes the form (we omit the $\delta$-function in Fourier space which reflects translation symmetry)
\be\label{G11}
G^{-1}=-\frac{\beta z^2_1}{2\Delta^2}B,
\ee
with $B$ an invertible $4\times 4$ matrix
\be\label{G12}
B=\left(\begin{array}{rrrr}
0&0&0&1\\
0&0&-1&0\\
0&-1&0&0\\
1&0&0&0
\end{array}\right)~,~B^2=1.
\ee
The propagator in momentum space is therefore given by 
\be\label{G13}
G=-\frac{2\Delta^2}{\beta}z^{-2}_1(q^2)B.
\ee
The non-vanishing correlation functions (\ref{Z6}) are
\be\label{G14}
F^{01}_{01}=F^{10}_{10}=-F^{10}_{01}=-F^{01}_{10}=-\frac{2\Delta^2}{\beta}z^{-2}_1(q^2).
\ee
This clearly differs from the observed structure of zweibein correlations
for our non-linear $\sigma$-model, as discussed in 
sect.~\ref{Zweibein correlations}.
It becomes clear however, that very different structures can also be 
accounted for by the ansatz (\ref{G3}). In appendix \ref{appe}
we discuss a possible form of
an effective action for scalars and zweibein that may be somewhat closer to 
our model.

\section{Zweibein correlations from scalar correlations}
\label{Zweibein correlations from scalar correlations}

For large $\beta$ one expects that the zweibein is well approximated by the scalar expectation values $\kl\psi\kr$, 
\be\label{ZS1}
e_\mu{^m}=2\Delta\kl\psi_\alpha\kr \sigma^m_{\alpha\beta}\partial_\mu\kl\psi_\beta\kr.
\ee
Similarly, one may try to approximate the zweibein 
fluctuations by the scalar fluctuations, 
$\delta\psi_\alpha=\psi_\alpha-\kl\psi_\alpha\kr$, 
by linearizing eq. \eqref{L2}, 
\be\label{ZS2}
f_\mu{^m}=\tilde e_\mu{^m}-e_\mu{^m}=2\Delta 
\{ \delta\psi_\alpha\sigma^m_{\alpha\beta}\partial_\mu\kl\psi_\beta\kr
+\kl\psi_\alpha\kr\sigma^m_{\alpha\beta}\partial_\mu\delta\psi_\beta\}.
\ee
In this approximation the zweibein correlations are approximated by the scalar correlations
\be\label{ZS3}
H_{\alpha\beta}(x,y)=\kl \delta\psi_\alpha(x)\delta\psi_\beta(y)\kr,
\ee
namely
\ba\label{ZS4}
&&F^{mn}_{\mu\nu}(x,y)=4\Delta^2\sigma^m_{\alpha\beta}\sigma^n_{\gamma\delta}
\Big\{\partial_\mu\kl\psi_\beta(x)\kr\partial_\nu\kl\psi_\delta(y)\kr\nn\\
&&\qquad -\kl\psi_\beta(x)\kr\partial_\nu\kl\psi_\delta (y)\kr
\frac{\partial}{\partial x^\mu}
-\partial_\mu\kl\psi_\beta(x)\kr\kl\psi_\delta(y)\kr
\frac{\partial}{\partial y^\nu}\nn\\
&&\qquad +\kl\psi_\beta(x)\kr\kl\psi_\delta(y)\kr
\frac{\partial}{\partial x^\mu}
\frac{\partial}{\partial y^\nu}\Big\}H_{\alpha\gamma}(x,y).
\ea

The relation \eqref{ZS4} between $F^{mn}_{\mu\nu}$ and $H_{\alpha\gamma}$ involves the expectation values of the scalar fields. We will evaluate them for the stripe configuration \eqref{F4}, \eqref{S1}. Since we are interested in large separations $(x-y)$ we employ the continuum limit. In the complex formulation we take for the cell averages continuous fields
\ba\label{ZS5}
&&\kl\varphi_1(x)\kr=\frac{1}{2\sqrt{2}}(1-i)\exp 
\left\{ -\frac{i\pi}{4\Delta}(x^0+x^1)\right\},\nn\\
&&\kl\varphi_2(x)\kr=\frac{1}{2\sqrt{2}}(1-i)\exp
\left\{-\frac{i\pi}{4\Delta}(-x^0+x^1)\right\}.
\ea
These fields are indeed invariant under the combined translations $t_0$ and $t_1$, cf. eqs. \eqref{S4}, \eqref{S5}. Taking partial derivatives of eq. \eqref{ZS5} reproduces the relations \eqref{S7} up to a factor $\pi/4$. This conversion factor for derivatives between the discrete and continuum formulation (e.g. discrete derivatives involving finite distances versus continuous derivatives, and cell averages versus continuous fields) has to be applied to eqs. \eqref{ZS1}, \eqref{ZS2}, \eqref{ZS4} if we use the standard partial derivatives, i.e. $\partial_\mu\to (4/\pi)\partial_\mu$. (The factor $4\Delta^2$ in eq. \eqref{ZS4} gets replaced by $(64/\pi^2)\Delta^2$ and the factor $2\Delta$ in eqs. \eqref{ZS1}, \eqref{ZS2} becomes $8\Delta/\pi$. With this replacement the evaluation of eq. \eqref{ZS1} for the stripe configuration \eqref{ZS5} yields indeed $e_\mu{^m}=\delta_\mu{^m}$.) We also observe the normalization
 $\kl\varphi^*_1\kr\kl\varphi_1\kr=\kl\varphi_2{^*}\kr\kl\varphi_2\kr=1/4$. 

We finally take into account that the normalization of the expectation value differs from eq. \eqref{I1} by multiplying $\kl\varphi_a\kr$ by a factor $Z_\varphi$, such that $|\kl\varphi_1\kr|^2+|\kl\varphi_2\kr|^2=Z^2_\varphi\leq 1$. In terms of the real fields $\kl\psi_\alpha\kr$ the stripe configuration becomes
\ba\label{ZS6}
\kl\psi_1\kr&=&\frac{Z_\varphi}{2\sqrt{2}}\Big[\cos \left(\frac{\pi}{4\Delta}(x^0+x^1)\right)-\sin 
\left(\frac{\pi}{4\Delta}(x^0+x^1)\right)\Big],\nn\\
\kl\psi_2\kr&=&\frac{Z_\varphi}{2\sqrt{2}}\Big[\cos\left(\frac{\pi}{4\Delta}(-x^0+x^1)\right)-\sin
\left(\frac{\pi}{4\Delta}(-x^0+x^1)\right)\Big],\nn\\
\kl\psi_3\kr&=&-\frac{Z_\varphi}{2\sqrt{2}}\Big[\sin\left(\frac{\pi}{4\Delta}(x^0+x^1)\right)+\cos 
\left(\frac{\pi}{4\Delta}(x^0+x^1)\right)\Big],\nn\\
\kl\psi_4\kr&=&-\frac{Z_\varphi}{2\sqrt{2}}\Big[\sin\left(\frac{\pi}{4\Delta}(-x^0+x^1)\right)+\cos
\left(\frac{\pi}{4\Delta}(-x^0+x^1)\right)\Big].\nn\\
\ea
Inserting eq. \eqref{ZS6} into eq. \eqref{ZS4} yields explicit expressions for the zweibein correlations as linear combinations of the scalar correlations. 
So far, we have not attempted to 
check this type of relations numerically.
The different qualitative behavior of scalar and zweibein 
correlations sheds doubts on the validity of 
such an approximation.

\section{Effective action for scalars and field equations}
\label{Field equations}

The quantum effective action $ \Gamma [ \psi ] $ for the scalar fields
is defined in the usual way by introducing sources for the scalar
fields in the functional integral, and performing a Legendre transform
of the generating functional for the connected Greens functions. It
includes all effects of fluctuations and generates the
one-particle-irreducible Greens functions. Thus the functions for an
arbitrary number of fields follow from $\Gamma$ by simple functional
differentiation. In this sense the knowledge of $ \Gamma $ amounts to a
solution of the model.
We do not attempt
here a computation of the effective action. We rather investigate a
simple ansatz which respects the symmetries of our model, namely
\be\label{E1}
\Gamma=\frac12\epsilon_{\mu\nu}\epsilon^{mn}\int d^2xV(\psi)\psi_\alpha\sigma^m_{\alpha\beta}\partial_\mu\psi_\beta\psi_\gamma\sigma^n_{\gamma\delta}\partial_\nu\psi_\delta,
\ee
where the $ \sigma^m_{\alpha\beta}$ matrices are defined in eq. (\ref{L3}). 
(We use $\psi_\alpha$ instead of $\kl \psi_\alpha(x)\kr$ in the following.) 
Eq. (\ref{E1}) equals the classical action \eqref{I2}  for $V(\psi)=-\beta$.
However,  we admit here a general ``scalar potential'' $V(\psi)$. If $V$ depends only on $\rho_1=\varphi^*_1\varphi_1=\psi^2_1+\psi^2_3$ and $\rho_2=\varphi^*_2\varphi_2=\psi^2_2+\psi^2_4$, with $V(\rho_1,\rho_2)=V(\rho_2,\rho_1)$, the effective action shares all symmetries of the classical action. 
We use the continuum version of the effective action (\ref{E1}) 
in order to demonstrate that the 
phase transition to the stripe phase also occurs in a continuum theory. 

The vacuum state (or thermal equilibrium state in case of a classical
statistical interpretation) is a solution of the quantum field
equations.  In our case, the quantum field equations for the scalar
fields are obtained from the functional derivative of the effective
action (\ref{E1}). In the absence of sources they read 
\ba\label{E2}
\frac{\delta\Gamma}{\delta\psi_\alpha}&=&\epsilon_{\mu\nu}\epsilon^{mn}
\{\psi_\gamma\sigma^n_{\gamma\delta}\partial_\nu\psi_\delta \nn\\
&\times& [2V\sigma^m_{\alpha\beta}\partial_\mu\psi_\beta+V'_\alpha\psi_\eta
\sigma^m_{\eta\beta}\partial_\mu\psi_\beta\psi_\alpha]\nn\\
&+&\partial_\mu[V\psi_\gamma\sigma^n_{\gamma\delta}\partial_\nu\psi_\delta]
\sigma^m_{\alpha\beta}\psi_\beta\}=0.
\ea
Here we use 
\be\label{E3}
\frac{\partial V}{\partial\psi_\alpha}=2V'_\alpha \psi_\alpha,
\ee
with $V'_\alpha=\partial V/\partial \rho_1$ for $\alpha=1,3$, $V'_\alpha=\partial V/\partial\rho_2$ for $\alpha=2,4$. The field equations always admit the solution $\psi_\alpha=0$ which corresponds to the disordered phase. We are interested here in the stripe solutions which are obtained for 
\be\label{E4}
2 V\sigma^m_{\alpha\beta}\partial_\mu\psi_\beta\epsilon^{\mu\nu}\epsilon_{mn}\hat e_\nu{^n}=-2 V'_\alpha\hat e\psi_\alpha,
\ee
with $V,V'_\alpha$ and 
\be\label{E5}
\hat e_\mu{^m}=\psi_\gamma\sigma^m_{\gamma\delta}\partial_\mu\psi_\delta~,~\hat e=\det (\hat e_\mu{^m}),
\ee
independent of $x$. Multiplication of eq. \eqref{E4} by $\psi_\alpha$ and summing over $\alpha$ yields as a condition for the existence of this type of solution 
\be\label{E6}
\rho_1\frac{\partial V}{\partial\rho_1}+\rho_2\frac{\partial V}{\partial\rho_2}=-2V.
\ee

We make the ansatz 
\be\label{E7}
\partial_\mu\psi_\beta=(U_\mu)_{\beta\gamma}\psi_\gamma, 
\ee
where only the elements $(1,3),(3,1),(2,4)$ and $(4,2)$ of the constant matrices $U_0$ and $U_1$ differ from zero. Eq. \eqref{E4} is obeyed for $\psi_\alpha\neq 0$ if 
\be\label{E8}
\epsilon_{\mu\nu}\epsilon^{mn}\hat e^n_\nu(2Vu^{\alpha,m}_\mu+V'_\alpha\hat e_\mu{^m})=0.
\ee
Here $u^{\alpha,m}_\mu$ is defined by the condition
\be\label{E9}
\sigma^m_{\alpha\beta}(U_\mu)_{\beta\gamma}=u^{\alpha,m}_\mu\delta_{\alpha\gamma}
\ee
and obeys
\be\label{E10}
\hat e_\mu\m=\sum^4_{\alpha=1}u^{\alpha,m}_\mu\psi^2_\alpha.
\ee
The squared matrices $(U_\mu)^2$ are diagonal
\be\label{E11}
(U^2_\mu)_{\alpha\beta}=\zeta_{\mu,\alpha}\delta_{\alpha\beta}~,~\zeta_{\mu,1}=\zeta_{\mu,3}~,~\zeta_{\mu,2}=\zeta_{\mu,4}.
\ee
Our ansatz \eqref{E7} requires then 
\be\label{E12}
(\partial_\mu)^2\psi_\alpha=\zeta_{\mu,\alpha}\psi_\alpha.
\ee
We will require the matrices $U_\mu$ to be antisymmetric, guaranteeing $\partial_\mu\rho_1=\partial_\mu\rho_2=0$. Then the coefficients $\zeta_{\mu,\alpha}$ are negative (or zero). As a consequence, one obtains solutions of eq. \eqref{E12} which are periodic in both $x^0$ and $x^1$. We typically will find solutions with $\zeta_{0,\alpha}=\zeta_{1,\alpha}$. They obey the wave equation 
\be\label{E13}
(\partial^2_0-\partial^2_1)\psi_\alpha=0.
\ee
It is remarkable how wave equations with two-dimensional Lorentz symmetry arise in a natural way from the field equations derived from the action \eqref{E1}. 

In analogy with eq. \eqref{ZS6} we  consider possible solutions 
of the type
\ba\label{E15}
\psi_1&=&c_1\Big[\cos\big(P_1(x^0+x^1)\big)-\sin \big(P_1(x^0+x^1)\big)\Big]\nn\\
\psi_3&=&-c_1\Big[\sin\big(P_1(x^0+x^1)\big)+\cos\big(P_1(x^0+x^1)\big)\Big]\nn\\
\psi_2&=&c_2\Big[\cos\big(P_2(-x^0+x^1)\big)-\sin\big(P_2(-x^0+x^1)\big)\Big]\nn\\
\psi_4&=&-c_2\Big[\sin\big(P_2(-x^0+x^1)\big)+\cos\big(P_2(-x^0+x^1)\big)\Big],\nn\\
\ea
with $\rho_1=2c^2_1,\rho_2=2c^2_2$ and antisymmetric matrices $U_\mu$ obeying
\ba\label{E16}
(U_0)_{13}&=&P_1~,~(U_1)_{13}=P_1,\nn\\
(U_0)_{24}&=&-P_2~,~(U_1)_{24}=P_2,\nn\\
\zeta_{0,1}&=&\zeta_{1,1}=-P^2_1~,~\zeta_{0,2}=\zeta_{1,2}=-P^2_2.
\ea
For $\alpha=1,3$ one finds for all $\mu$ and $m$ that $u^{\alpha,m}_\mu=P_1$ whereas for $\alpha=2,4$ one has $u^{\alpha,m}_\mu=P_2$ if $\mu=m$, and $u^{\alpha,m}_\mu=-P_2$ if $\mu\neq m$. Eq. \eqref{E10} yields
\be\label{E17}
\hat e_\mu\m=\left(\begin{array}{ccc}
P_1\rho_1+P_2\rho_2&,&P_1\rho_1-P_2\rho_2\\
P_1\rho_1-P_2\rho_2&,&P_1\rho_1+P_2\rho_2
\end{array}\right).
\ee
For the particular stripe solution \eqref{ZS6} with $P_1=P_2=\pi/(4\Delta)~,~\rho_1=\rho_2=Z^2_\varphi/4$ one recovers
\be\label{E18}
\hat e^m_\mu=\frac{Z^2_\varphi}{8\Delta}\delta^m_\mu~,~e^m_\mu=\frac{8\Delta}{\pi}\hat e_\mu\m
=Z^2_\varphi\delta^m_\mu.
\ee

For general $P_1,P_2,\rho_1,\rho_2$ one has 
\be\label{E19}
\hat e=\det(\hat e_\mu\m)=4P_2P_2\rho_1\rho_2,
\ee
and eq. \eqref{E8} is obeyed for 
\be\label{E20}
\rho_1\frac{\partial V}{\partial {\rho_1}}=\rho_2\frac{\partial V}{\partial \rho_2}=-V.
\ee
If eq. \eqref{E20} has a solution for suitable values of $\rho_1$ and $\rho_2$ we therefore find solutions with arbitrary $P_1$ and $P_2$. 
The condition (\ref{E20}) implies the condition (\ref{E6}).
We conclude that for a potential which admits a solution of eq. (\ref{E20})
stripe solutions (\ref{E15}) exsits with arbitrary ``momenta''
$ P_1$ and $P_2$.

We may try to interprete the effective action (\ref{E1}) as an approximation
to the continuum limit of the quantum effective action 
which corresponds to the microscopic lattice action (\ref{I2}).
In this case one expects a dependence of the 
shape of $V$ on the parameter $\beta$. (This extends to 
a parameter dependence of $V$ for other models in the same 
universality class.) A phase transition from the disordered 
phase with $ \psi_\alpha = 0 $ to the stripe phase occurs at $ \beta_c $ 
if for $ \beta > \beta _c $ eq. (\ref{E20}) has a solution and 
if for the corresponding stripe solution (\ref{E15})
the action (\ref{E1}) becomes negative. 
At the phase transition for $\beta= \beta_c$ the 
effective action in the stripe 
phase vanishes, such that the 
free energy $ \Gamma $ has the same value for the disordered and the stripe
phase. A first order transition is realized if for $ \beta=\beta_c$ 
the stripe solution still has a nonvanishing ``order parameter'' $ \psi_\alpha$.

In the remainder of this appendix we discuss 
simple shapes of the potential $V$ that realize
the first order phase transition that we observe in our numerical results.
A constant potential, e.g. $V=-\beta$, does not admit solutions with $\psi_\alpha\neq 0$. This has a simple explanation: the action is then a pure quartic polynomial of $\psi$, such that for any value of $\psi$ for which $\Gamma<0$ the rescaled field $(1+\epsilon)\psi, \epsilon>0$, leads to an even smaller value of $\Gamma$, thus excluding an extremum for $\Gamma\neq 0$. For the microscopic action this problem is cured by the non-linear constraint \eqref{I1}, which would be translated to the continuum language as $\rho_1+\rho_2=1/2$. 

Solutions of the condition (\ref{E20}) exist for a wide class of 
non-trivial potentials $ V(\rho_1 , \rho_2 ) $ without
invoking constraints for $\rho_1 $ and $ \rho_2 $. As a first example 
we consider a potential $V(\rho),\rho=\rho_1+\rho_2$. The stripe solutions correspond then to an extremum of the combination 
\be\label{E21}
W(\rho)=\rho^2 V(\rho)~,~\frac{\partial W}{\partial \rho}(\rho_0)=0.
\ee
Indeed, eq. (\ref{E21}) implies that the condition (\ref{E20}) has a solution
with $\rho_1=\rho_2=\rho_0/2$. (For $\beta\to\infty$ one would expect $\rho_0\to 1/2$.) The value of the effective action for stripe
 solutions with $\rho_1=\rho_2=\rho_0/2$ is given by $W_0=W(\rho_0)$,
\be\label{E22}
\Gamma_0=\int d^2 x\hat e V(\rho_0)=P_1P_2\int d^2 xW_0.
\ee
For $P_1=P_2=\pi/(4\Delta)$ the action per lattice point equals $(\pi^2/8)W_0$, such that for $\beta\to\infty$ one expects $W_0\to-(8/\pi^2)\beta$. We observe that $\Gamma_0$ in eq. \eqref{E22} can be made arbitrarily negative for $P_1P_2\to\infty$. This ``ultraviolet divergence'' is cut off by the lattice regularization. We may consider $(\pi/4\Delta)$ as the maximal momentum, say in the $x^1$-direction.

Many different forms of the potential $V(\rho_1,\rho_2)$ are conceivable. For example, if $V$ only depends on the combination $\rho_1\rho_2$ the effective action \eqref{E1} is invariant under local Lorentz transformations
\be\label{E23}
\psi'_{1,3}(x)=e^{\alpha(x)/2}\psi_{1,3}(x)~,~\psi'_{2,4}(x)=e^{-\alpha(x)/2}\psi_{2,4}(x)
\ee
for which the zweibein transforms as
\ba\label{E24}
(\hat e^0_\mu)'&=&\cosh \alpha~\hat e^0_\mu+\sinh \alpha~\hat e^1_\mu,\nn\\
(\hat e^1)'&=&\cosh \alpha~\hat e^1_\mu+\sinh \alpha~\hat e^0_\mu.
\ea

For the example (with positive constants $a,b$)
\be\label{E25}
V=-a+b\rho_1\rho_2
\ee
one has $\rho_1\partial V\partial\rho_1=\rho_2\partial V/\partial\rho_2=b\rho_1\rho_2$ and eq. \eqref{E20} is met for 
\be\label{E26}
\rho_{1,0}\rho_{2,0}=\frac{a}{2b},
\ee
with 
\be\label{E27}
V(\rho_{10},\rho_{20})=-\frac a2.
\ee
Indeed, the combination $\hat e V$ takes for the configurations \eqref{E15} the form 
\be\label{E28}
\hat e V=4P_1P_2[-a\rho_1\rho_2+b(\rho_1\rho_2)^2].
\ee
For $P_1P_2>0$ this has a minimum for eq. \eqref{E26} with $\hat e>0,(\hat e V)_0=-P_1P_2 a^3/(2b^2)$. We observe a degeneracy of the minimum under global Lorentz transformations \eqref{E23}. 

For $P_1P_2<0$ the combination $\hat eV$ has a maximum with $\hat e<0,\hat e V_0>0$. The action can become arbitrarily negative for $P_1P_2<0$ and large $\rho_1\rho_2$. We may prevent this to happen by imposing a Lorentz invariant constraint
\be\label{E29}
\rho_1\rho_2<\frac{a}{b}.
\ee
An interesting issue concerns the possibility to use eq. \eqref{E1} with eq. \eqref{E25} for the microscopic action $S$, and to replace the non-linear constraint \eqref{I1} by the condition  \eqref{E29}. This would permit to realize Lorentz symmetry as an exact symmetry of the model. 

For our model the observed first order transition at $\beta_c$ indicates an effective potential that is more complicated than the form \eqref{E25}. Indeed, a first order transition is described if a possible local minimum of $\hat e V$ with $\rho_1,\rho_2$ different from zero occurs for $\beta_c$ at $(\hat eV)_0=0$, while for $\beta<\beta_c$ one has $(\hat eV)_0>0$. For $\beta<\beta_c$ the absolute minimum of the effective action \eqref{E1} will then be found at $\psi=0$, corresponding to the disordered phase.

As an example, consider
\be\label{E30}
V=-a+b\rho_1\rho_2-\frac c2(\rho_1+\rho_2)+\frac d4(\rho_1+\rho_2)^2.
\ee
Possible solutions of eq. \eqref{E20} with $\rho_1=\rho_2=\rho/2$ occur for 
\be\label{E31}
\rho_0=\frac{3c\pm\sqrt{9c^2+32a(d+b)}}{4(d+b)}
\ee
with 
\be\label{E32}
V_0=V(\rho_0)=-\frac{a}{2}-\frac18c\rho_0.
\ee
For $d+b>0$ and $c>0$ one finds a critical value 
\be\label{E33}
a_c=-\frac{c^2}{4(d+b)},
\ee
such that for $a>a_c$ one has $V_0<0$, and for $a<a_c$ the solution \eqref{E31} with smallest $V$ occurs for $V_0>0$. Thus for $a<a_c$ the disordered phase with $\psi=0$ is realized and we may associate the critical value $\beta_c$ with $a(\beta_c)=a_c$. Indeed, for $\rho_1=\rho_2=\rho/2$ we can consider
\be\label{E34}
W(\rho)=\rho^2V(\rho)=-a\rho^2-\frac c2\rho^3+\frac14(d+b)\rho^4.
\ee
For $a<0$ one has a local minimum of $W(\rho)$ at $\rho=0$. A second local minimum exists for $c>(4/3)\sqrt{-2a(d+b)}$. For $c>2\sqrt{-a(d+b)}$, corresponding to $a>a_c$, this second minimum occurs for negative $W$ and is therefore deeper than the minimum at $\rho=0$. At the critical value $a_c$ the order parameter $\sim \rho_0$ jumps from zero to
\be\label{E35}
\rho_{0,c}=\frac{2c}{d+b}.
\ee

Having found a satisfactory description of the first order phase
transition from the disordered phase to the stripe phase one may ask
if a suitable shape of $V$ can also account for the scalar correlation
functions in the continuum limit. In principle, the inverse scalar
correlation functions can be obtained from the second variation of the
effective action, evaluated for the appropriate solution of the field
equation \eqref{E2}. We have not yet performed a computation of the 
correlation functions that correspond to the effective action (\ref{E1}).

\section{Effective action for zweibein and scalars}
\label{appe}

The effective action for scalars and zweibein can be defined by introducing appropriate sources
\ba\label{A1}
W[\eta,t]&=&\ln\int {\cal D}\tilde \psi\exp \{-S+\sum_{\tilde y} t^\mu_m(\tilde y)\tilde e^m_\mu(\tilde y)\nn\\
&&+\sum_{\tilde z}\eta_\alpha(\tilde z)\tilde\psi_\alpha(\tilde z)\},\nn\\
\frac{\partial W}{\partial t^\mu_m(y)}&=&\kl\tilde e_\mu\m(\tilde y)\kr=e_\mu\m(\tilde y),\nn\\
\frac{\partial W}{\partial \eta_\alpha(\tilde z)}&=&\kl\tilde \psi_\alpha(\tilde z)\kr =\psi_\alpha(\tilde z),
\ea
and performing a Legendre transform
\be\label{A2}
\Gamma[\psi,e]=-W+\sum_{\tilde y}t_m{^\mu}(\tilde y)e_\mu\m(\tilde y)+
\sum_{\tilde z}\eta_\alpha(\tilde z)\psi_\alpha(\tilde z).
\ee
This yields the exact quantum field equations
\be\label{A3}
\frac{\partial\Gamma}{\partial e_\mu\m(\tilde y)}=t_m{^\mu}(\tilde y)~,~
\frac{\partial\Gamma}{\partial\psi_\alpha(\tilde z)}=\eta_\alpha (\tilde z).
\ee
The scalar effective action $\Gamma[\psi]$ discussed in the preceding section is obtained for $t_m{^\mu}(\tilde y)=0$. It can be inferred from $\Gamma[\psi,e]$ by solving the field equation $\partial\Gamma/\partial e_\mu\m(\tilde y)=0$ with solution $e^{(0)m}_\mu(y)[\psi]$ being a functional of $\psi$. Then $\Gamma[\psi]=\Gamma\big[\psi,e^{(0}[\psi]\big]$. 

Due to lattice diffeomorphism invariance of the action the continuum
limit of the effective action is invariant under general coordinate
transformations\cite{LDI}. 
Besides diffeomorphism symmetry the effective action also
preserves the discrete reflection symmetries of the lattice action as
well as the continuous flavor symmetry.

Let us try an ansatz for the continuum limit of the effective action which is consistent with the symmetries
\ba\label{A4}
\Gamma&=&\frac12\epsilon_{\mu\nu}\epsilon^{mn}\int d^2 x\big \{ \psi_\alpha\sigma^m_{\alpha\beta}\partial_\mu\psi_\beta 
\big[V_1(\psi)\psi_\gamma\sigma^n_{\gamma\delta}\partial_\nu\psi_\delta\nn\\
&&+V_2(\psi)e_\nu{^n}\big]+V_3(\psi)e_\mu\m e_\nu\m\Big\}.
\ea
The field equation for the zweibein reads
\be\label{A5}
\frac12 \epsilon_{\mu\nu}\epsilon^{mn}(2V_3e_\nu\m+V_2\psi_\gamma\sigma^n_{\gamma\delta}\partial_\nu\psi_\delta)
=t_m{^\mu}.
\ee
For $V_{2,3}\neq 0$, and in the absence of sources $t_m{^\mu}=0$, the solution is
\be\label{A6}
e_\nu{^n}=-\frac{V_2}{2V_3}\psi_\gamma\sigma^n_{\gamma\delta}\partial_\nu\psi_\delta.
\ee
The proportionality between $ e_\nu^n$ and $\psi_\gamma \sigma_{\gamma\delta}^n
\partial_\nu \psi_\delta $ is realized in our model for 
$ \beta > \beta_c $, as can be seen in Fig.~\ref{bein_disconnected}. The proportionality
factor $ N^{(e)} / N^{(e\varphi)} $ corresponds to 
$ - V_2 / 2 V_3 $, evaluated for the appropriate $ \rho_1 \rho_2$
 (cf. appendix \ref{Field equations}), and taking the 
proper normalization of partial derivatives into account.

Insertion of eq. (\ref{A6}) into eq. \eqref{A4} yields eq. \eqref{E1} with 
\be\label{A7}
V=V_1-\frac{V^2_2}{4V_3}.
\ee
The field equation for the scalars \eqref{A3} for $\eta_\alpha=0$ yield, after insertion of the solution \eqref{A6}, precisely eq. \eqref{E2} with $V$ given by eq. \eqref{A7}. We can therefore take over the discussion of the preceding section. In particular, for $\beta>\beta_c$ one finds the wave solutions \eqref{E15}, with 
\be\label{A8}
\psi_\alpha\sigma^m_{\alpha\beta}\partial_\mu\psi_\beta\sim\delta^m_\mu~,~e_\mu\m\sim\delta^m_\mu.
\ee

Eq. \eqref{A6} explains our finding $e_\mu\m\sim\delta^m_\mu$ for all $\beta>\beta_c$, and that $e_\mu\m$ as well as $\psi$ vanish simultaneously for $\beta<\beta_c$. We realize a flat space geometry without tuning of parameters for all $\beta>\beta_c$.

\end{appendix}

\end{document}